# Comparative analysis of physical properties of some binary transition metal carbides XC (X = Nb, Ta, Ti): Insights from a comprehensive ab-initio study

Razu Ahmed, Md. Mahamudujjaman, Md. Asif Afzal, Md. Sajidul Islam, R.S. Islam, S.H. Naqib*
Department of physics, University of Rajshahi, Rajshahi 6205, Bangladesh
*Corresponding author email: salehnaqib@yahoo.com

## Abstract

Binary metallic carbides belong to a technologically prominent class of materials. We have explored the structural, mechanical, electronic, optical, and some thermophysical properties of XC (X = Nb, Ta, Ti) binary metallic carbides in details employing density functional theory based first-principles method. Some of the results obtained are novel. Study of elastic constants and moduli shows that XC (X = Nb. Ta, Ti) compounds possess low level of elastic anisotropy, reasonably good machinability, mixed bonding characteristics with ionic and covalent contributions, brittle nature and high Vickers hardness with very high Debye temperatures. The mechanical and dynamical stability conditions are fulfilled. The bulk modulus and Young's modulus of TiC are lower than those of NbC and TaC. All the XC (X = Nb, Ta, Ti) compounds are hard suitable for heavy duty engineering applications. The electronic band structures with finite electronic energy density of states at the Fermi level reveal metallic character of XC (X = Nb, Ta, Ti). Presence of both covalent and ionic bondings are also evident from the charge density distribution maps of XC (X = Nb, Ta, Ti) compounds. The vibrational properties such as phonon dispersion curves and phonon density of states for XC (X = Nb, Ta, Ti) compounds are also calculated. Positive phonon frequencies at the $\Gamma$-point suggest that the solids under investigation are dynamically stable and capable of efficient thermal transport. The optical parameters are found to be almost isotropic. The optical absorption, reflectivity spectra, and the static index of refractive of XC (X = Nb, Ta, Ti) show that the compounds hold promise to be used in optoelectronic device sectors. Debye temperature, melting temperature, lattice thermal conductivity, and minimum phonon thermal conductivity of the compounds under study are high and show excellent correspondence with the elastic and bonding characteristics. Extremely high melting temperature of TaC indicates that this compound is a good candidate material for high-temperature applications.

## 1. Introduction

Transition metal carbides (TMCs) are a large and rich group of industrially relevant compounds with diverse set of physical properties [1]. The combination of metals with light covalent-bond forming carbon atoms often leads to materials with many useful physical properties, such as extreme stiffness, wear resistance, corrosion resistance, high melting point, high hardness, good thermal and electrical conductivity, and even superconductivity [1–11]. Examples for binary transition metal carbides with remarkable properties are the extremely high melting points of HfC (4201 K [12]) and TaC (4223 K [12]). The compound TaC has a comparatively high superconducting transition temperature $T_c$ = 10.3 K [12]. Hence, it is of interest to understand their structure-property relations. These compounds have traditionally been utilized in drill bits and rocket nozzles because of their high strength and durability under conditions of high pressure and temperature [13]. Due to high hardness, these materials have found use in snow tires, golf shoe spikes, and cutting tools [13]. In ferrous alloys, they are the components that give steel its toughness [13]. However, they also have interesting optical, electronic, and magnetic properties and have been used for electrical contacts, diffusion barriers, and other uses [13]. TMCs can also be employed in low-friction coatings where the TMC phase is enclosed inside a matrix of amorphous carbon [14,15].

Groups 4 and 5 TMCs all produce monocarbides with a face-centered-cubic (fcc) B1 crystal structure, including TiC, ZrC, HfC, VC, NbC, and TaC [16]. These carbides have two interpenetrating metal and carbon fcc lattices, which are similar to the NaCl structure. The (100) and (111) surfaces are the most important crystalline planes. The coexistence of metal and carbon atoms is what distinguishes the surface (100). The metal and carbon atom layers in the (111) orientation alternate, thus the (111) surface can be terminated by either metal or carbon atom [16].

In this study, three binary TMC materials, XC (X = Nb, Ta, Ti) are explored using the Kohn-Sham density functional theory (KS-DFT). There are some previous theoretical and experimental works on our chosen materials [14,17–26]. In those articles, structural, elastic, electronic, optical dielectric properties, chemical bonding properties, and charge density distributions have been studied with varying depth. Among these, elastic properties have been explored lightly, and to the best of our knowledge, a detailed study of elastic properties including Cauchy pressure, tetragonal shear modulus, Kleinman parameter, machinability index, Pugh's index, anisotropy in elastic moduli are still lacking. A number of important thermo-mechanical properties including the minimum phonon thermal conductivity, thermal conductivities along different crystallographic directions, and wavelength of the dominant phonon modes in XC (X = Nb, Ta, Ti) have not yet been discussed theoretically and experimentally at all. Furthermore, the mechanical anisotropy characterized by the directional dependence of Young's modulus, shear modulus, linear compressibility, and Poisson's ratio of TaC and TiC have not yet been discussed at all, except for NbC [27]. Detailed study of bonding properties is yet to be done. The optical properties of TaC have not been explored comprehensively. A detailed understanding of a compound's mechanical response to external stress is crucial for any potential use. The ductile/brittle behavior is closely associated with machinability; elastic anisotropy indices give idea about the possible mechanical failure modes. Optical properties are important to know to select a material for optoelectronic device applications. Thermo-physical parameters provide with knowledge regarding the behavior of a solid at different temperatures. Furthermore, it is of significant interest to understand the variations in the physical properties of XC (X = Nb, Ta, Ti) with different metal atoms. For a thorough comprehension of the elastic, bonding, electronic, optical and thermo-physical properties, we have performed a comparative study of the physical properties of XC (X = Nb, Ta, Ti) compounds. The results obtained are compared with those found in previous studies where available. The electronic properties, such as electronic band structure, density of states, Fermi surface topology and charge density distribution are related to charge transport, optical and electronic thermal processes. Mulliken bond population analysis elucidates the bonding nature of the compounds under study. All the mechanical properties are related to the bonding characteristics.

The rest of the paper is organized in the following manner: In Section 2, we have discussed the computational methodology. In Section 3, we have presented the computational results, and analyzed those. Finally, the major findings of our calculations are discussed and summarized in Section 4.

## 2. Computational methodology

Density functional theory (DFT) is the most widely employed formalism for *ab-initio* calculations in crystalline solids where the ground state of a crystalline system is found by solving the Kohn-Sham equation [28] with periodic boundary conditions (involving Bloch states). In this work, DFT based CAmbridge Serial Total Energy Package (CASTEP) code [29] has been used to explore various physical properties of the titled compounds. This code implements the total energy plane-wave pseudopotential method. In this study, we have used local density approximation (LDA), generalized gradient approximation (GGA) of the Perdew–Burke–Ernzerhof (PBE), GGA of the PW91, and GGA of the PBEsol exchange-correlation functionals. Considering the experimental structural parameters, LDA [30,31] for TaC, GGA (PBE) [32] for NbC and GGA (PW91) [33,34] for TiC provide the results for the ground state crystal geometry. The importance of proper geometry optimization is significant. Therefore, we have presented results for the functionals which yield the best structural parameters for XC (X = Nb, Ta, Ti) in the subsequent sections. Ultrasoft Vanderbilt-type pseudopotentials were used to calculate the electron-ion interactions. This scheme relaxes the norm-conserving constraint but produces a smooth and computationally friendly pseudopotential that saves computational time without affecting the accuracy of the obtained parameters appreciably [35].

The valence electron configurations of Nb, Ta, Ti, and C have been taken as [$4p^6 4d^4 5s^1$], [$5d^3 6s^2$], [$3p^6 3d^2 3s^2$], and [$2s^2 2p^2$], respectively. In this paper, Monkhorst-Pack scheme is used for the *k*-point sampling in the first irreducible Brillouin zone (BZ). The selected *k*-point meshes have the sizes of 6 × 6 × 6 for NbC and TaC, 5 × 5 × 5 for TiC. The cut-off energy for the plane wave expansion has been taken as 550 eV for NbC and TiC, and 350 eV for TaC. On the other hand, to obtain a smooth Fermi-surfaces of XC (X = Nb, Ta, Ti) compounds, 24 × 24 × 24 *k*-points mesh have been used. For optimizing geometry via minimizing the total energy and the internal forces, the BFGS (Broyden-Fletcher- Goldfarb-Shanno) algorithm [36] is chosen. The structure is relaxed up to a convergence threshold of $5\times10^{-6}$ eV-atom$^{-1}$ for energy, 0.01 eV Å$^{-1}$ for the maximum force, 0.02 GPa for maximum stress and $5\times10^{-4}$ Å for maximum displacement.

Based on the stress-strain approach [37], the single crystal elastic constants, $C_{ij}$, are computed. There are three independent elastic constants in a cubic crystal ($C_{11}$, $C_{12}$ and $C_{44}$). Using the Voigte-Reusse-Hill (VRH) method [38,39], all the other polycrystalline elastic parameters, including the bulk modulus (B), shear modulus (G), and Young modulus (Y) can be evaluated from the values of the single crystal elastic constants $C_{ij}$. Using the optimized crystal structures of XC (X = Nb, Ta, Ti), the electronic band structure, total and atom resolved partial density of states (TDOS and PDOS, respectively) are obtained.

The complex dielectric function, $\varepsilon(\omega) = \varepsilon_1(\omega) + i\varepsilon_2(\omega)$ can be used to calculate all the energy/frequency dependent optical parameters. The real part $\varepsilon_1(\omega)$ of dielectric function $\varepsilon(\omega)$ has been obtained from the imaginary part $\varepsilon_2(\omega)$ using the Kramers-Kronig relationships. The imaginary part, $\varepsilon_2(\omega)$ of the complex dielectric function has been estimated within the momentum representation of matrix elements of transition between occupied and unoccupied electronic states in the valence and conduction states, respectively. The CASTEP supported formula giving the imaginary part of the dielectric function is given below:

$$\varepsilon_2(\omega) = \frac{2e^2\pi}{\Omega\varepsilon_0} \sum_{k,v,c} |\langle \psi_k^c | \hat{u} \cdot \vec{r} | \psi_k^v \rangle|^2 \delta(E_k^c - E_k^v - E) \tag{1}$$

where, $\Omega$ is the volume of the unit cell, $\omega$ is angular frequency (or equivalently energy) of the incident electromagnetic wave (photon), $\hat{u}$ is the unit vector defining the polarization direction of the incident electric field, $e$ is the electronic charge, $\psi_k^c$ and $\psi_k^v$ are the conduction and valence band wave functions at a fixed wave-vector $k$, respectively. During the optical transition, the conservation of energy and momentum is implemented by the delta function. All optical properties, including refractive index, optical conductivity, reflectivity, absorption coefficient, and energy loss function, can be estimated from the dielectric function $\varepsilon(\omega)$, once it is known at different energies [40].

The average sound velocity in the target compounds is used to compute the Debye temperature. The other thermo-physical parameters are calculated from the crystal density, elastic constants, and elastic moduli of XC (X = Nb, Ta, and Ti).

A widely used method for understanding a material's bonding properties is the Mulliken bond population study [41]. The projection of the plane-wave states onto a linear combination of atomic orbital (LCAO) basis sets [42,43] has been used for XC (X = Nb, Ta, Ti) compounds. The Mulliken bond population analysis can be implemented using the Mulliken density operator written on the atomic (or quasi-atomic) basis as follows:

$$P_{\mu'v}^M(g) = \sum_{g'} \sum_{v'} P_{\mu v'}(g') S_{v'v}(g - g') = L^{-1} \sum_k e^{-ikg} (P_k S_k)_{\mu v'} \tag{2}$$

and the net charge on an atomic species A is defined as,

$$Q_A = Z_A - \sum_{\mu \in A} P_{\mu\mu}^m(o) \tag{3}$$

where $Z_A$ represents the charge of the nucleus or atomic core (used in simulations of the atomic pseudopotential).

## 3. Results and analysis

### *3.1. Structural properties*

The crystal structure of XC (X = Nb, Ta, Ti) compounds is cubic (NaCl-type) with space group $Fm\bar{3}m$ (no. 225) [3,17,24]. The schematic crystal structure of XC (X = Nb, Ta, Ti) compounds is shown below in Fig. 1. Each unit cell of XC (X = Nb, Ta, Ti) contains four formula units and 8 atoms (4 X and 4 C) in total. The atomic positions in a unit cell are: X atoms at (0, 0, 0) and C atoms at (0.5, 0.5, 0.5). In Table 1, the results of first-principles calculations of structural properties of these materials along with available theoretical and experimental lattice parameters are listed [3,17,24]. Calculated lattice parameters show good agreement with theoretical and experimental results. The lattice parameters of NbC are slightly higher than that of TaC and TiC. This variation can be largely accounted due to the variation in the atomic radii of Ti, Nb, and Ta.

The phase stability of a compound is determined by the formation enthalpy; the negative value of formation enthalpy implies chemical stability. The formation enthalpies of the compounds under study have been calculated using the following equation:

$$\Delta H_f(XC) = E_{total}(XC) - E(X) - E(C) \qquad (4)$$

where, $E_{total}(XC)$ is the total enthalpy of XC (X = Nb, Ta, Ti) compounds per formula unit; $E(X)$ is the total enthalpy of the X atom (X = Nb, Ta, Ti) and $E(C)$ is the total enthalpy of the C atom. The lower the formation enthalpy, the more stable the compound. It is worth noting that, the experimental structural parameters are determined at room temperature while the *ab-initio* calculations are performed assuming absolute zero temperature.

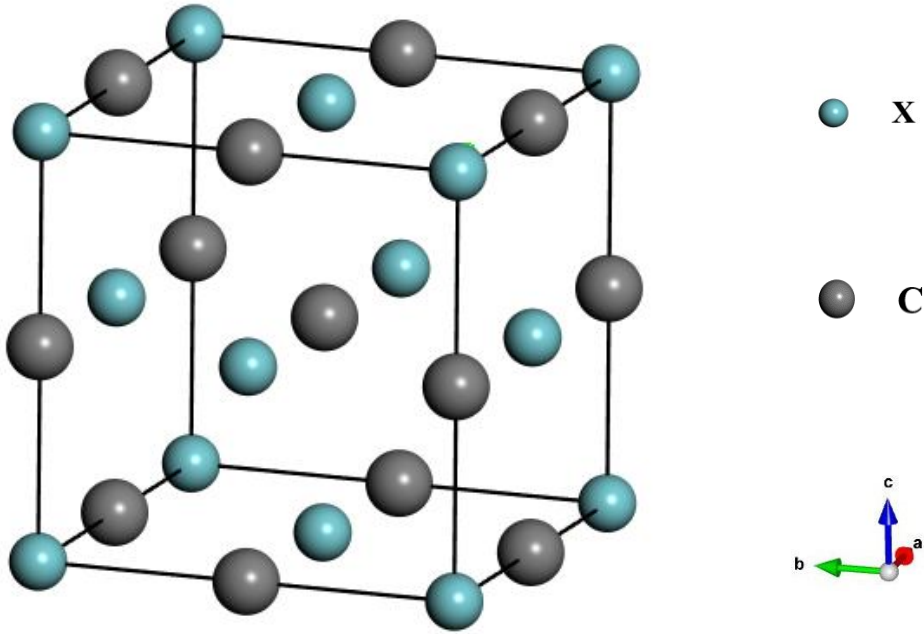

**Fig. 1.** Schematic crystal structure of XC (X = Nb, Ta, Ti) compounds. The crystallographic directions are shown.

**Table 1.** Calculated, theoretical, and experimental lattice constants $a$ (Å), equilibrium volume $V_o$ (Å$^3$), and the formation enthalpy (KJ/mole) of the XC (X = Nb, Ta, Ti) compounds.

| Compound | $a$ | $V_0$ | Formation enthalpy | Ref. |
|---|---|---|---|---|
| NbC | 4.4786 | 89.834 | -224.60 | This work |
| | 4.4676 | 89.171 | ---- | [24][Expt.] |
| | ---- | ---- | -141.00 | [25][Expt.] |
| | 4.480 | 90.189 | -140.90 | [17][Theo.] |
| TaC | 4.4262 | 86.713 | -239.66 | This work |
| | 4.4557 | 88.460 | ---- | [24][Expt.] |
| | 4.580 | 95.284 | -142.80 | [17][Theo.] |
| TiC | 4.330 | 81.204 | -201.23 | This work |
| | 4.328 | 81.030 | ---- | [3][Expt.] |
| | ---- | ---- | -184.00 | [25][Expt.] |
| | 4.330 | 81.390 | -180.90 | [17][Theo.] |

The estimated enthalpies are somewhat greater in magnitude than those values obtained before. In case of TiC, the agreement is quite good.

## 3.2. Elastic properties

The elastic constants of a crystalline solid provide the link between mechanical and dynamical behavior of the compound and determine the response of the crystal to external forces. The bulk mechanical responses to applied stress are characterized by the bulk modulus ($B$), shear modulus ($G$), Young's modulus ($Y$), and Poisson's ratio ($\sigma$). These parameters also give important information concerning the nature of chemical bonding in solids. For the materials with cubic symmetry, there are only three independent single-crystal elastic constants: $C_{11}$, $C_{12}$, and $C_{44}$. A stress-strain approach based on the generalized Hooke's law is used to calculate elastic constants and the results are disclosed in Table 2. It can be seen that the calculated values in this work are consistent with other experimental and theoretical results. According to Born-Huang conditions, a cubic crystal system needs to satisfy the following criteria for mechanical stability [44]:

$$C_{11} > 0,\ C_{44} > 0,\ C_{11}\text{-}C_{12} > 0,\ C_{11}+2C_{12} > 0 \tag{5}$$

All the elastic constants of XC (X = Nb, Ta, Ti) satisfy the mechanical stability criteria. This implies that they are mechanically stable. We observe that, $C_{44}$, which reflects the resistance to shear deformation, is lower than $C_{11}$, which is related to the unidirectional strain along the principle crystallographic directions. This means that the cubic cell is more easily deformed by a shear in comparison to the unidirectional stress. The tetragonal shear modulus, $C'$, which is a measure of stiffness, is related to $C_{11}$ and $C_{12}$ by the following relation: $C' = (C_{11}\text{-}C_{12})/2$. A positive value of this parameter suggests dynamical stability of the solid. The bulk modulus, shear modulus, Young's modulus, and Poisson's ratio can be estimated using the widely used Voigt-Reuss-Hill (VRH) approximation. Voigt approximation gives us the upper limit of the polycrystalline elastic moduli and Reuss approximation gives us the lower limit of the polycrystalline elastic moduli. The real value lies between Voigt and Reuss bounds. Hill later proposed the arithmetic average of the two limits, which closely represents the practical situation. To get the bulk modulus ($B$) and shear modulus ($G$) (by Voigt-Reuss-Hill (VRH) method), Young's modulus ($Y$), Poisson's ratio ($\sigma$), and shear anisotropy factor ($A$) at zero pressure, the following formulae have been used [44–49]:

$$B_H = \frac{B_V + B_R}{2} \tag{6}$$

$$G_H = \frac{G_V + G_R}{2} \tag{7}$$

$$Y = \frac{9BG}{(3B + G)} \tag{8}$$

$$\sigma = \frac{(3B - 2G)}{2(3B + G)} \tag{9}$$

$$A = \frac{2C_{44}}{(C_{11}-C_{12})} \tag{10}$$

The results are presented in Table 3, which are in good agreement with previous reports. Isotropic shear modulus and bulk modulus are the gross measures of the bonding strength of a material. From Table 3, the larger value of $B$ compared to $G$ indicates that the mechanical stabilities of XC (X = Nb, Ta, Ti) are expected to be controlled by the shearing strain [50]. The covalent nature of a material increases with the increasing value of Young's modulus and it is a measure of the stiffness (resistance) of an elastic material to a change in its length [51,52]. The bulk modulus and Young's modulus of TiC are lower than those of NbC and TaC, respectively. The lower value of Young's modulus of TiC compared to that of NbC and TaC also indicates that TiC is less stiff than NbC and TaC. A number of thermo-mechanical parameters are linked to the elastic moduli. For example, the lattice thermal conductivity ($K_L$) and Young's modulus of a material are related as: $K_L \sim \sqrt{Y}$ [53].

Some of the factors on which the brittle and ductile natures of a given system depend are Pugh's ratio ($B/G$), Poisson's ratio ($\sigma$), and Cauchy pressure ($C''$). Pugh [54–56] proposed an empirical relationship to distinguish the mechanical properties (ductility and brittleness) solids. If the value of Pugh's ratio is greater than 1.75, the material is predicted to be ductile; otherwise, the material is expected to be brittle. In our case, all the materials are expected to show brittle behavior. As shown in Table 3, the Pugh's ratio of NbC is lower than that of TiC and TaC, respectively, which indicates that NbC is more brittle than TiC and TaC. Poisson's ratio ($\sigma$) is a measure of materials deformation (expansion or contraction) along the perpendicular direction of loading. It also gives a measure of the stability of solids against shear. The numerical limit for Poisson's ratio of a material is, $-1.0 \leq \sigma \leq 0.5$ [44]. If $\sigma = 0.5$, no volume change occurs during elastic deformation [57]. The values of $\sigma$ are much lower than 0.5 for XC (X = Nb, Ta, Ti) compounds, implying a large volume change under stress. Poisson's ratio is an important parameter that determines the various mechanical properties of crystalline solids. It can predict the ductility or brittleness of materials with the critical value of 0.26. If $\sigma$ is less (greater) than 0.26, the material is brittle (ductile) [58,59]. Based on this criterion, we can conclude that XC (X = Nb, Ta, Ti) compounds are brittle. This is consistent with the result obtained from the Pugh's ratio. The nature of interatomic forces in solids can also be predicted by the value of $\sigma$ [60,61]. If $\sigma$ remains between 0.25 and 0.50, central force interaction will dominate. Otherwise, non-central force will dominate. Thus, in XC (X = Nb, Ta, Ti) compounds, non-central force should dominate the atomic bonding. In completely ionic bonded solids, the value of $\sigma$ is approximately 0.33, while it is around 0.10 in purely covalent bonded crystal [62]. This implies that ionic bonding should be dominant in TaC and TiC, while covalent bonding should be dominating in NbC (Table 3). The overall bonding characters of these binary metal carbides under consideration should exhibit mixed character with different proportions of covalent and ionic contributions. Furthermore, Poisson's ratio signifies the level of plasticity of a solid against shear. Larger the Poisson's ratio better is the plasticity. The Cauchy pressure, ($C''$) is another useful mechanical parameter for solids. The Cauchy pressure of a material is defined as $C'' = (C_{12} - C_{44})$. A ductile material has positive Cauchy pressure, whereas a brittle material has negative Cauchy pressure [63]. Cauchy pressure is also used to describe the angular characteristics of atomic bonding in a material [64]. The presence of ionic and covalent bonding in a material is related with positive and negative values of the Cauchy pressure, respectively. The Pettifor's rule [64] states that a material with large positive Cauchy pressure has significant metallic bonds and exhibits high level of ductility. On the other hand, a material with negative Cauchy pressure possesses more angular bonds and thus exhibits more brittleness and significant covalent bonding. As a result, negative value of Cauchy pressure predicts that XC (X = Nb, Ta, Ti) compounds are brittle in nature and some angular (covalent) bonding are present in XC (X = Nb, Ta, Ti).

A number of useful mechanical performance indicators, namely the machinability index ($\mu_M$) [65], Kleinman parameter ($\zeta$) and hardness parameters ($H_{micro}$ [66,67], $H_{macro}$ [68], $(H_v)_{Tian}$ [69], $(H_v)_{Teter}$ [70], and $(H_v)_{Mazhnik}$ [71]) are calculated using the following widely employed equations.

$$\mu_M = \frac{B}{C_{44}} \tag{11}$$

$$\zeta = \frac{C_{11} + 8C_{12}}{7C_{11} + C_{12}} \tag{12}$$

$$H_{micro} = \frac{(1-2\sigma)Y}{6(1+\sigma)} \tag{13}$$

$$H_{macro} = 2[\left(\frac{G}{B}\right)^2 G]^{0.585} - 3 \tag{14}$$

$$(H_V)_{Tian} = 0.92(G/B)^{1.137} G^{0.708} \tag{15}$$

$$(H_V)_{Teter} = 0.151 G \tag{16}$$

$$(H_V)_{Mazhnik} = \gamma_0 \chi(\sigma) Y \tag{17}$$

In Equation (17), $\chi(\sigma)$ is a function of Poisson's ratio and can be evaluated from:

$$\chi(\sigma) = \frac{1 - 8.5\sigma + 19.5\sigma^2}{1 - 7.5\sigma + 12.2\sigma^2 + 19.6\sigma^3}$$

where $\gamma_0$ is a dimensionless constant with a value of 0.096.

The ease with which a material can be machined using cutting/shaping tools is known as machinability and it is measured by a parameter known as machinability index. Information regarding machinability of a material is becoming valuable in today's industry, because it defines the optimum level of machine utilization, cutting forces, temperature and plastic strain. It can also be used as a measure of plasticity and dry lubricating nature of a solid [72–75]. A high value of $\mu_M$ indicates excellent lubricating properties, i.e., lower friction. The value of $\mu_M$ in XC (X = Nb, Ta, Ti) compounds implies a good level of machinability. The machinability index of TaC is higher than that of NbC and TiC respectively. These materials with appreciable covalent and ionic bondings show high machinability. This is an attractive feature for machine tools applications.

Generally, the value of Kleinman parameter lies between 0 and 1. According to Kleinman [76], the lower limit of $\zeta$, represents significant contribution of bond stretching or contracting to resist external stress whereas the upper limit corresponds to significant contribution of bond bending to resist external load. From Table 3, it is evident that, mechanical strengths in XC (X = Nb, Ta, Ti) compounds are mainly derived from the bond stretching or contracting contribution.

Information regarding the hardness of a solid is essential to understand its elastic and plastic properties. This is also essential from the applications point of view. Among the elastic constants and moduli, $C_{44}$ and $G$ are considered as the best hardness predictor of solids [77,78]. The obtained values of $H_{micro}$, $H_{macro}$, $(H_V)_{Tian}$, $(H_V)_{Teter}$, and $(H_V)_{Mazhnik}$ of XC (X = Nb, Ta, Ti) compounds are found to be consistent with the predictions based on the values of $C_{44}$ and $G$ and are disclosed in Table 4. However, $H_{micro}$ is observed to be higher than that of other hardness indicators obtained using the Chen's formula. The difference in the hardness values comes due to the parameters involved in the equations [(13) – (17)]. It is found that the hardness of TaC is greater than that of NbC and TiC. It appears that TaC possesses high level of hardness. At present, to the best of our knowledge, there is no previous study on hardness to compare the present results.

One of the most significant issues with the surface hard coatings on heavy-duty equipment is the development of cracks, especially in metals and ceramic materials. Fracture toughness, $K_{IC}$ is a parameter, which can evaluate the resistance of a material to crack/fracture initiation. The formula for $K_{IC}$ of a material is as follows [79]:

$$K_{IC} = \alpha_0^{-1/2} V_0^{1/6} [\xi(\sigma)Y]^{3/2} \qquad (18)$$

where $V_0$ = volume per atom; $\alpha_0$ = 8840 GPa for covalent and ionic crystals; $\xi(\sigma)$ is a dimensionless parameter and a is a function of Poisson's ratio ($\sigma$), which can be found from:

$$\xi(\sigma) = \frac{1 - 13.7\sigma + 48.6\sigma^2}{1 - 15.2\sigma + 70.2\sigma^2 - 81.5\sigma^3}$$

As presented in Table 4, the values of $K_{IC}$ of the XC (X = Nb, Ta, Ti) compounds are 6.69, 8.66, and 4.85 MPam$^{1/2}$, respectively. It is found that the highest resistance to surface crack formation and propagation is offered by the TaC compound and it is found to be consistent with the value of hardness.

**Table 2**. Calculated elastic constants, $C_{ij}$ (GPa) and tetragonal shear modulus, $C'$ (GPa) for XC (X = Nb, Ta, Ti) compounds.

| Compound | $C_{11}$ | $C_{12}$ | $C_{44}$ | ($C_{11}$ - $C_{12}$) | $C'$ | Ref. |
|---|---|---|---|---|---|---|
| NbC | 720.14 | 92.40 | 184.98 | 627.74 | 313.87 | This work |
|  | 557.30 | 162.40 | 146.50 | 394.90 | --- | [17] |
| TaC | 899.04 | 101.88 | 201.67 | 797.16 | 398.58 | This work |
|  | 562.00 | 159.20 | 146.40 | 402.80 | --- | [17] |
| TiC | 513.09 | 120.00 | 171.05 | 393.09 | 196.55 | This work |
|  | 523.20 | 115.50 | 206.90 | 407.70 | --- | [17] |

**Table 3.** Calculated elastic moduli (all in GPa), Pugh's ratio, Poisson's ratio, Cauchy pressure (in GPa), machinability index ($\mu_M$), Kleinman parameter ($\zeta$) for the XC (X = Nb, Ta, Ti) compounds.

| Compound | B | G | Y | B/G | $\sigma$ | C″ | $\mu_M$ | $\zeta$ | Ref. |
|---|---|---|---|---|---|---|---|---|---|
| NbC | 301.65 | 228.94 | 548.14 | 1.32 | 0.197 | -92.58 | 1.63 | 0.28 | This work |
|  | 294.00 | 164.60 | 483.90 | 1.79 | 0.226 | --- | --- | --- | [17] |
| TaC | 367.59 | 265.88 | 642.69 | 1.38 | 0.208 | -99.79 | 1.82 | 0.26 | This work |
|  | 293.50 | 166.40 | 491.80 | 1.76 | 0.221 | --- | --- | --- | [17] |
| TiC | 251.03 | 180.83 | 437.45 | 1.39 | 0.210 | -51.05 | 1.47 | 0.38 | This work |
|  | 251.40 | 205.70 | 481.40 | 1.22 | 0.181 | --- | --- | --- | [17] |

**Table 4.** Calculated hardness (GPa) and fracture toughness (MPam$^{1/2}$) based on elastic moduli and Poisson's ratio of XC (X = Nb, Ta, Ti) compounds.

| Compound | $H_{micro}$ | $H_{macro}$ | $(H_V)_{Tian}$ | $(H_V)_{Teter}$ | $(H_V)_{Mazhnik}$ | $K_{IC}$ |
|---|---|---|---|---|---|---|
| NbC | 46.22 | 31.78 | 31.50 | 34.57 | 29.47 | 6.69 |
| TaC | 51.34 | 32.88 | 33.15 | 40.15 | 32.08 | 8.66 |
| TiC | 34.95 | 25.50 | 25.12 | 27.31 | 21.84 | 4.85 |

The bulk modulus along *a*-, *b*- and *c*- axis (known as bulk modulus under uniaxial strain or directional bulk modulus) and anisotropies of the bulk modulus can be defined as [80]:

$$B_a = a\frac{dP}{da} = \frac{\Lambda}{1+\alpha+\beta} \qquad (19)$$

$$B_b = a\frac{dP}{db} = \frac{B_a}{\alpha} \qquad (20)$$

$$B_c = c\frac{dP}{dc} = \frac{B_a}{\beta} \qquad (21)$$

$$A_{B_a} = \frac{B_a}{B_b} = \alpha \qquad (22)$$

$$A_{B_c} = \frac{B_c}{B_b} = \frac{\alpha}{\beta} \qquad (23)$$

where, $\Lambda = C_{11} + 2C_{12}\alpha + C_{22}\alpha^2 + 2C_{13}\beta + C_{33}\beta^2 + 2C_{33}\alpha\beta$ and for cubic crystals, $\alpha = \beta = 1$. $A_{B_a}$ and $A_{B_c}$ represent anisotropies of bulk modulus along *a*-axis and *c*-axis with respect to *b*-axis, respectively. A value of $A_{B_a} = A_{B_c} = 1$ indicates isotropy. This means that directional bulk modulus is isotropic in XC (X = Nb, Ta, Ti) compounds. Along different crystallographic axes, estimated directional bulk moduli for XC (X = Nb, Ta, Ti) compounds are equal and presented in Table 5. These values are higher than those of the isotropic polycrystalline bulk modulus. This arises from the fact that the pressure in a state of uniaxial strain for a given crystal density generally differs from the pressure in a state of hydrostatic stress at the same density of the solid [80].

Elastic anisotropy has influence on the development of plastic deformation in crystals, formation and propagation of microscale cracks in ceramics, plastic relaxation in the thin films etc. Therefore, it is quite important to evaluate the elastic anisotropy factors for solids to predict their behavior under different conditions of external stresses [81–83]. The elastic/mechanical anisotropy indices of XC (X = Nb, Ta, Ti) compounds are investigated in this section. The following widely used relations are used to calculate anisotropy factors:

The shear anisotropic factor for a cubic crystal can be quantified by three factors [80,84]: For {100} shear planes between the ⟨011⟩ and ⟨010⟩ directions, the shear anisotropic factor, $A_1$ is,

$$A_1 = \frac{4C_{44}}{C_{11}+C_{33}-2C_{13}} \qquad (24)$$

For the {010} shear plane between ⟨101⟩ and ⟨001⟩ directions the shear anisotropic factor, $A_2$ is,

$$A_2 = \frac{4C_{55}}{C_{22}+C_{33}-2C_{23}} \qquad (25)$$

For the {001} shear planes between ⟨110⟩ and ⟨010⟩ directions, the anisotropic factor, $A_3$ is,

$$A_3 = \frac{4C_{66}}{C_{11}+C_{22}-2C_{12}} \qquad (26)$$

The calculated values of these anisotropic factors are enlisted in Table 5. The calculated values of $A_1$, $A_2$, and $A_3$ are different from 1. These anisotropy factors imply that XC (X = Nb, Ta, Ti) compounds are moderately anisotropic with respect to shearing stress along different crystal planes and all the components of the shear anisotropic factors are equal (reflecting cubic symmetry). This anisotropy arises from the nature of the bond forming atomic orbitals in different crystal planes and directions. From Table 5, it is seen that the anisotropic behavior of TaC is greater than that of NbC and TiC, respectively. Zener anisotropy factor, $A$ has been calculated using following equation [45]:

$$A = \frac{2C_{44}}{C_{11}-C_{12}} \qquad (27)$$

Zener anisotropy factor shows same value as shear anisotropy factor of XC (X = Nb, Ta, Ti) compounds.

The universal log-Euclidean index is defined by using a log-Euclidean formula [85,86]:

$$A^L = \sqrt{[\ln(\frac{B_V}{B_R})]^2 + 5[\ln(\frac{C_{44}^V}{C_{44}^R})]^2} \qquad (28)$$

In this scheme, the Voigt and Reuss values of $C_{44}$ is obtained from [87]:

$$C_{44}^V = C_{44}^R + \frac{3}{5}\frac{(C_{11}-C_{12}-2C_{44})^2}{3(C_{11}-C_{12})+4C_{44}} \qquad (29)$$

and

$$C_{44}^R = \frac{5}{3}\frac{C_{44}(C_{11}-C_{12})}{3(C_{11}-C_{12})+4C_{44}} \qquad (30)$$

Kube and Jong [85,86] suggested that the value of $A^L$ lies in the range $0 \leq A^L \leq 10.26$ for inorganic crystalline compounds and 90% of these compounds have $A^L < 1$. For perfect isotropy, $A^L = 0$. Based on the value of $A^L$, it is difficult to ascertain whether a material is layered/lamellar or not. But, the majority (78%) of these inorganic crystalline compounds with high $A^L$ value exhibit layered/lamellar structure [87]. Compounds with higher $A^L$ values show strong layered structural features and with lower $A^L$ values show non layered structure. From the comparatively low value of $A^L$ we can predict that TaC and NbC, exhibit moderately layered type of configuration. But for the very low value of $A^L$, TiC is expected to exhibit non layered feature and possesses very small level of anisotropy in particular.

The universal anisotropy index, $(A^U, d_E)$, equivalent Zener anisotropy measure, $A^{eq}$, percentage anisotropy in compressibility, $A_B$ and anisotropy in shear, $A_G$ (or $A_C$) for crystals are calculated using following standard equations [50,85,88,89]:

$$A^U = 5\frac{G_V}{G_R} + \frac{B_V}{B_R} - 6 \geq 0 \qquad (31)$$

$$d_E = \sqrt{A^U + 6} \qquad (32)$$

$$A^{eq} = \left(1 + \frac{5}{12}A^U\right) + \sqrt{(1+\frac{5}{12}A^U)^2 - 1} \qquad (33)$$

$$A_B = \frac{B_V - B_R}{B_V + B_R} \qquad (34)$$

$$A_G = \frac{G_V - G_R}{2G_H} \qquad (35)$$

Ranganathan and Ostoja Starzewski [88] have defined the universal anisotropy index, which provides a singular measure of bulk anisotropy in solids. $A^U$ is called *universal* because of its applicability to all sorts of crystal symmetries. The condition for an isotropic crystal is $A^U = 0$. Derivation from this value, which must be positive, suggests the presence of anisotropy. $A^U$ for XC (X = Nb, Ta, Ti) compounds, shows same trend as $A^L$. For $A^{eq}$, a value of 1 represents isotropy, while any other value indicates anisotropy. The calculated values of $A^{eq}$ for XC (X = Nb, Ta, Ti) compounds are given in Table 5 predicting that XC (X = Nb, Ta, Ti) compounds are moderately anisotropic. For $A_B$ and $A_G$, zero value represents elastic isotropy and a value of 1 corresponds to the highest anisotropy. However, Chung and Buessem [90] refrained from extending $A_G$ to crystals with lower symmetries because, in addition to the shear modulus, the bulk modulus influences the anisotropy of crystals having other than cubic symmetry [90]. It is notable that compared to other measures, $A_G$ predicts a much lower value of anisotropy index. Since $A_B$ is the percentage of the measurement of anisotropy in compressibility, a zero value of $A_B$ for XC (X = Nb, Ta, Ti) compounds with cubic structure proves that bulk modulus has no influence on the anisotropic elastic and mechanical properties.

**Table 5.** Shear anisotropic factors ($A_1$, $A_2$ and $A_3$), Zener anisotropy factor $A$, universal log-Euclidean index $A^L$, the universal anisotropy index ($A^U$, $d_E$), equivalent Zener anisotropy measure $A^{eq}$, anisotropy in shear $A_G$, anisotropy in compressibility $A_B$, anisotropy in bulk modulus, and directional bulk modulus (in GPa) for XC (X = Nb, Ta, Ti) compounds.

| Compound | $A_1$ | $A_2$ | $A_3$ | $A$ | $A^L$ | $A^U$ | $d_E$ | $A^{eq}$ | $A_G$ | $A_B$ | $A_{B_a}$ | $A_{B_c}$ | $B_a$ |
|---|---|---|---|---|---|---|---|---|---|---|---|---|---|
| NbC | 0.59 | 0.59 | 0.59 | 0.59 | 0.42 | 0.34 | 2.52 | 1.69 | 0.03 | 0 | 1 | 1 | 1323.43 |
| TaC | 0.51 | 0.51 | 0.51 | 0.51 | 0.67 | 0.58 | 2.57 | 1.98 | 0.05 | 0 | 1 | 1 | 1634.24 |
| TiC | 0.87 | 0.87 | 0.87 | 0.87 | 0.03 | 0.02 | 2.45 | 1.14 | 0.002 | 0 | 1 | 1 | 1015.15 |

Three-dimensional (3D) direction dependent Young modulus, linear compressibility (inverse of bulk modulus), shear modulus, and Poisson's ratio should all have spherical shapes for a mechanically isotropic solid, however any deviation from spherical shape indicates anisotropy. We have shown ELATE [91] generated 2D and 3D plots of directional dependence of Young modulus, shear modulus, linear compressibility, and Poisson's ratio for XC (X = Nb, Ta, Ti) compounds here. As can be seen from Figs. 2 – 5, there is small deviation from spherical shape in the 3D figures of Y, G, and σ signifying some degree of anisotropy. But there is no deviation from spherical shape in the 3D figure of β signifying complete isotropy. The ELATE generated plots also demonstrate that the projections of direction dependent Y, β, G, and σ in the *ab*-plane are circular. This implies that the elastic properties are isotropic within the basal plane is very small. The maximum and minimum values of Y, β, G, and σ and their maximum to minimum ratios are presented in Table 6. These ratios are useful indicators of elastic anisotropy.

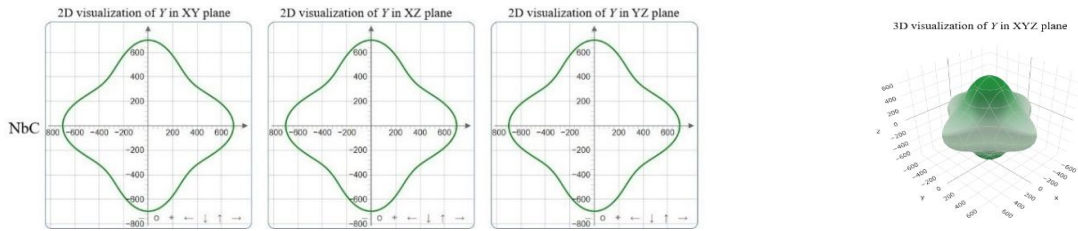

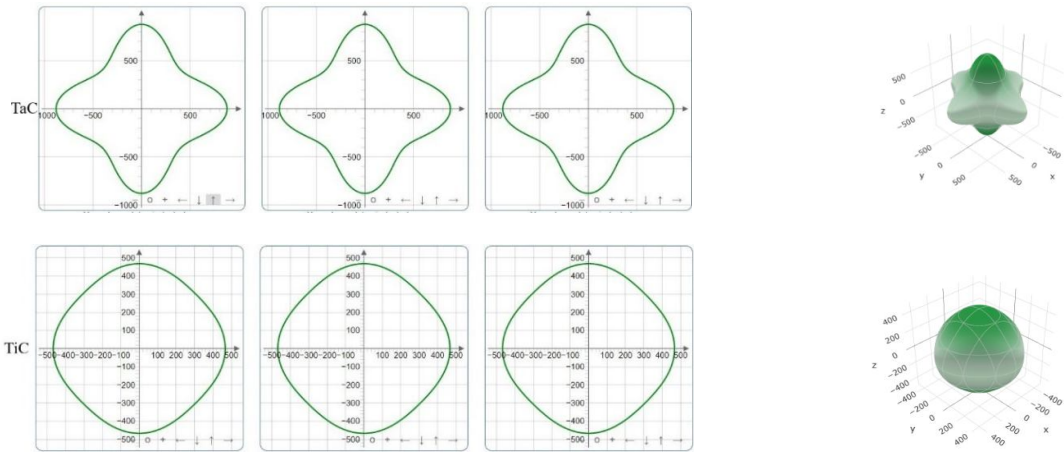

**Fig. 2.** Directional variation in Young's modulus (*Y*) of XC (X = Nb, Ta, Ti) compounds.

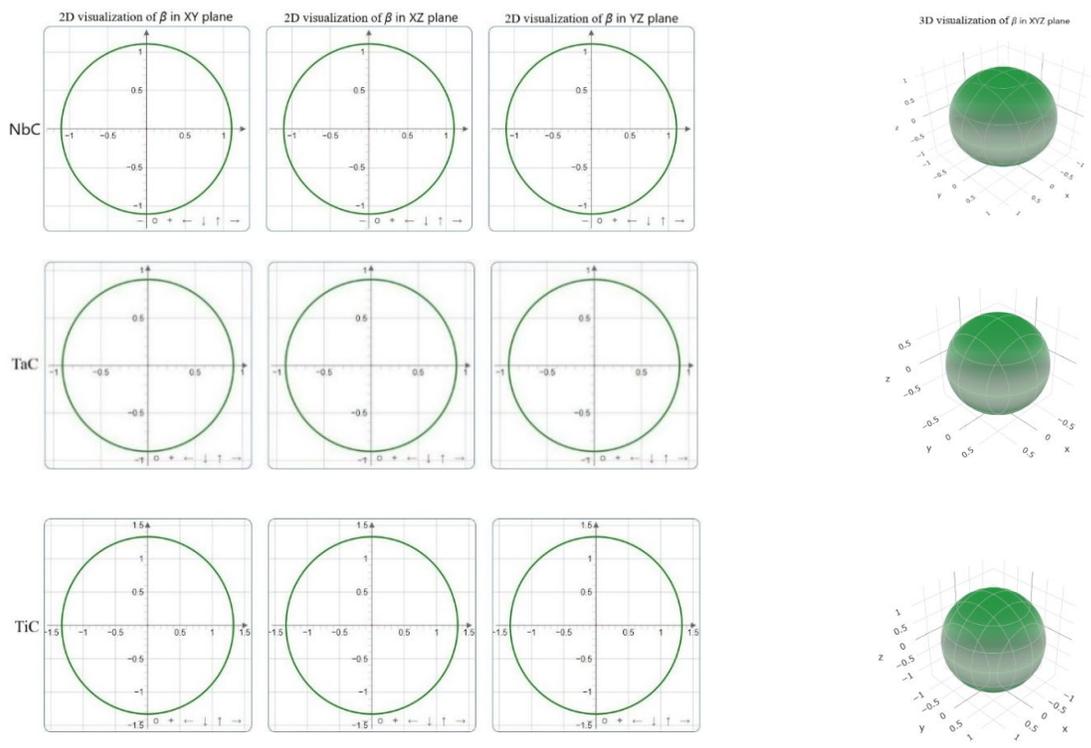

**Fig. 3.** Directional variation in linear compressibility (*β*) of XC (X = Nb, Ta, Ti) compounds

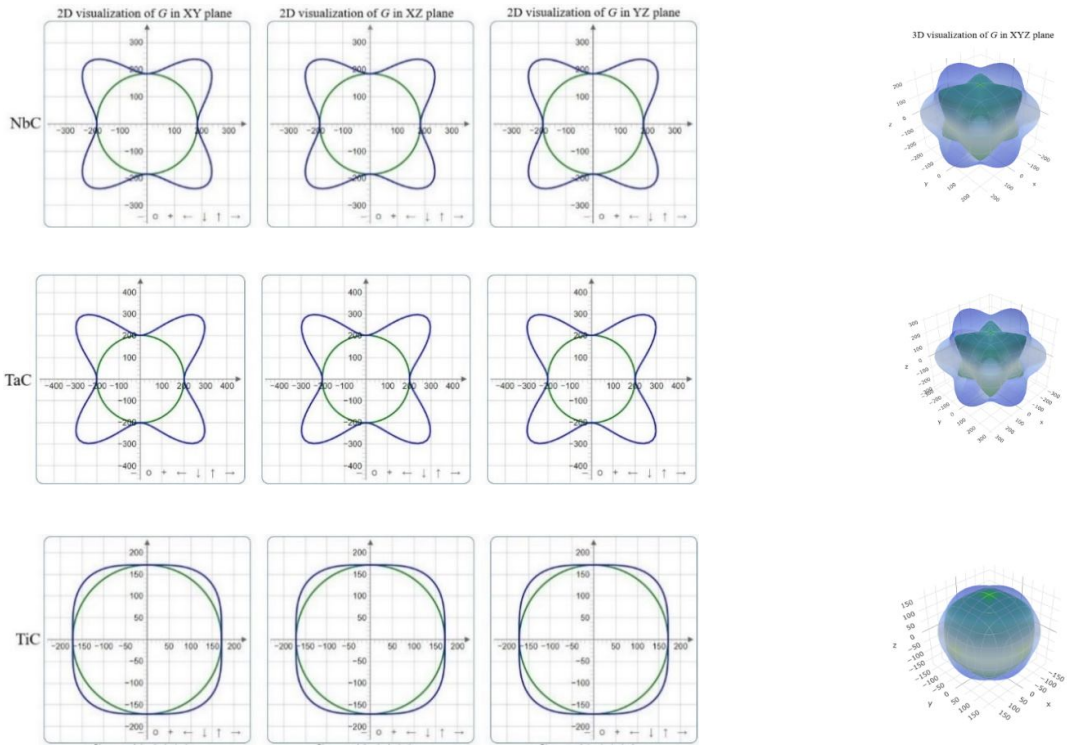

**Fig. 4.** Directional variation in shear modulus ($G$) of XC (X = Nb, Ta, Ti) compounds.

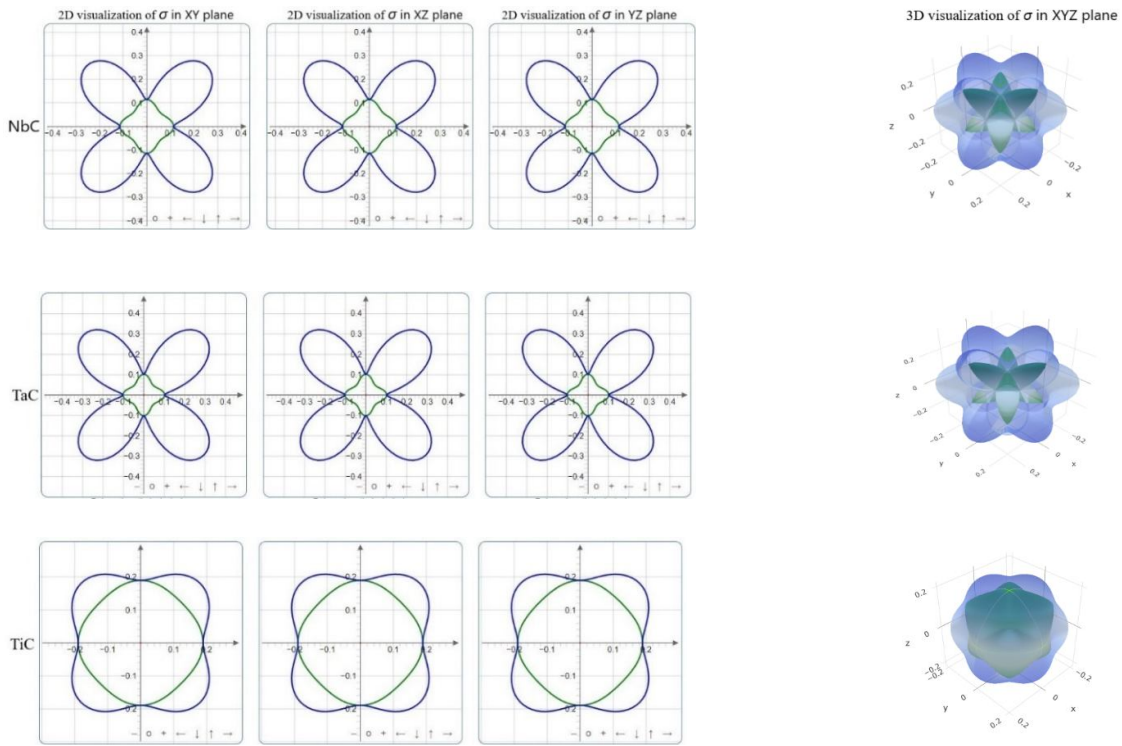

**Fig. 5.** Directional variation in Poisson's ratio ($\sigma$) of XC (X = Nb, Ta, Ti) compounds.

**Table 6.** The minimum and maximum values of Young's modulus (GPa), compressibility (TPa$^{-1}$), shear modulus (GPa), Poisson's ratio, and their ratios for XC (X = Nb, Ta, Ti).

| Compound | $Y_{min}$ | $Y_{max}$ | $A_Y$ | $\beta_{min}$ | $\beta_{max}$ | $A_\beta$ | $G_{min}$ | $G_{max}$ | $A_G$ | $\sigma_{min}$ | $\sigma_{max}$ | $A_\sigma$ |
|---|---|---|---|---|---|---|---|---|---|---|---|---|
| NbC | 460.77 | 699.13 | 1.52 | 1.105 | 1.105 | 1.00 | 184.98 | 313.87 | 1.69 | 0.08 | 0.36 | 4.41 |
| TaC | 511.47 | 878.31 | 1.72 | 0.91 | 0.91 | 1.00 | 201.67 | 398.58 | 1.98 | 0.07 | 0.42 | 6.29 |
| TiC | 418.17 | 467.60 | 1.12 | 1.33 | 1.33 | 1.00 | 171.05 | 196.54 | 1.15 | 0.17 | 0.26 | 1.47 |

### *3.3. Phonon dynamics*

The characteristics of phonons have fundamental importance for studying crystalline materials. From the phonon dispersion spectra (PDS) and phonon density of states (PHDOS), many properties of a material can be measured directly or indirectly [92]. Phonon dispersion curves impart the information about the dynamical stability, phase transition and contribution of vibrations in thermal properties of a material [93]. Besides, all the charge transport and thermo-mechanical properties are influenced by phonon dynamics. The phonon dispersion curves and phonon density of states of XC (X = Nb, Ta, Ti) compounds at absolute zero have been calculated to check the dynamical stability. This computation has been carried out applying density functional perturbation theory (DFPT) based finite displacement method (FDM) [94,95]. Fig. 6 shows the phonon dispersion spectra and phonon density of states of titled materials along high symmetry direction of the Brillouin zone (BZ) at zero pressure and temperature. We have put phonon dispersion curve and phonon density of states graphs side by side in order to compare the bands and their corresponding density of states. The absence of negative frequencies in the phonon dispersion curves encompassing the entire Brillouin zone strongly indicates the dynamical stability of the compounds of interest. In the low frequency region, there are three branches. These branches represent acoustic modes. These modes are divided into transverse acoustic (TA) and longitudinal acoustic (LA) modes. On the other hand, the branches in the high frequency are optical modes. Generally, the acoustic modes with low frequencies arise due to the vibration of heavy atom whereas the optical modes with high frequencies come from the vibration of light atom. The zero frequency of acoustic modes at the $\Gamma$ point is another proof of dynamical stability of XC (X = Nb, Ta, Ti) compounds.

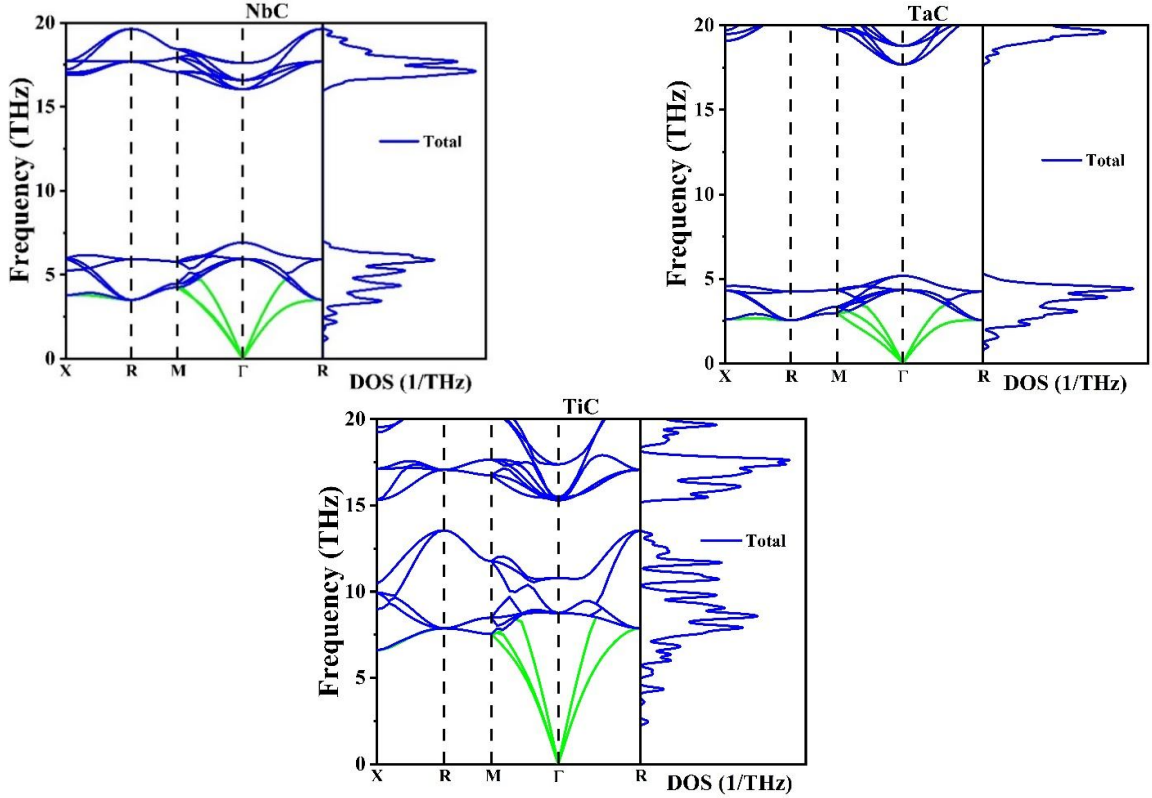

**Fig. 6.** Calculated phonon dispersion spectra and phonon DOS for XC (X = Nb, Ta, Ti) compounds at zero pressure. The green lines show the phonon branches around the center of the BZ.

## *3.4. Electronic band structure and density of states*

Electronic band structures for optimized crystal structure of XC (X = Nb, Ta, Ti) along several high symmetry directions (*X-R-M-Γ-R*) in the first Brillouin zone are depicted in Fig. 7. The horizontal broken line shows the Fermi level. The total number of bands for NbC, TaC, and TiC are 84, 58, and 83, respectively. The band structure features clearly disclose that a number of bands with varying degree of dispersion cross the Fermi level. This confirms the metallic character of XC (X = Nb, Ta, Ti) compounds. Bands crossing the Fermi level are indicated in colored lines. The band numbers that cross the Fermi level are 33, 34, 35, 36, 37, and 38 for NbC, 17, 18, 19, 20, and 21 for TaC and 30, 31, 32, 33, and 34 for TiC. Highly dispersive nature of the bands crossing the Fermi level near *Γ*-point for XC (X = Nb, Ta, Ti) compounds are suggestive of high mobility of the charge carriers in this region of the BZ. It is interesting to note that the bands crossing near the *Γ*-point display both electron- and hole-like features for XC (X = Nb, Ta, Ti). Highly dispersive bands imply low charge carrier effective mass [96–98] and high charge mobility. Overall, for all three compounds under investigation, the curves along *R-M*, directions are less dispersive, which indicates relatively high effective mass of charge carriers and consequently low mobility in these directions. On the other hand, the band curves along *M-Γ* and *Γ-R* directions show high dispersion, which indicates low effective mass and high mobility of electrons.

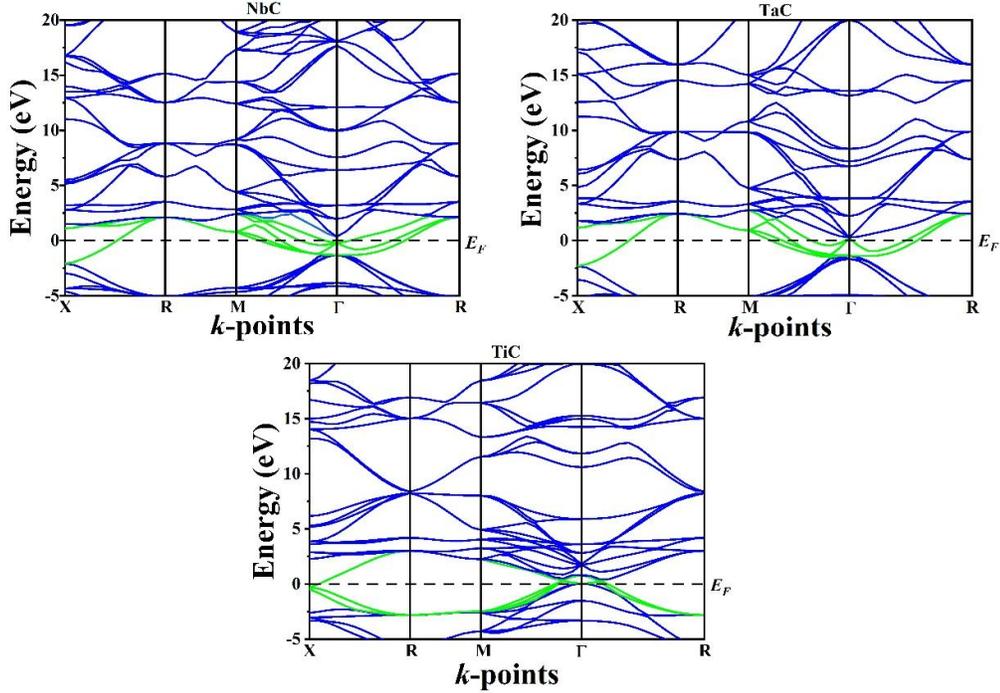

**Fig. 7.** The electronic band structures of XC (X = Nb, Ta, Ti) compounds along the high symmetry directions of the *k*-space within the first Brillouin zone.

The calculated total and partial density of states (TDOSs and PDOSs, respectively) of XC (X = Nb, Ta, Ti) compounds, as a function of energy, ($E$-$E_F$), are illustrated in Fig. 8. The vertical dashed line at 0 eV represents the Fermi level, $E_F$. To understand the contribution of each atom to the TDOSs, we have calculated the PDOSs of Nb, Ta, Ti, and C atoms in XC. The non-zero values of TDOSs at the Fermi level is an evidence that NbC, TaC and TiC should exhibit metallic electrical conductivity. At the Fermi energy, the values of TDOSs for NbC, TaC, and TiC are 1.87, 1.60, and 1.35 states per eV respectively. Hence, NbC should be the most conducting among the three. Near the Fermi level the main contribution on TDOSs comes from Nb-4*d* orbitals for NbC, Ta-5*d* for TaC, and Ti-3*d* for TiC. Thus, these electronic states should dominate the electrical conductivity of the XC compounds. The chemical and mechanical stability of XC compounds is also mainly affected by the properties of Nb-4*d*, Ta-5*d*, and Ti-3*d* bonding electronic states. There is a significant overlap in energy between the C-2*p* and Nb-4*d* bands in NbC, C-2*p* and Ta-5*d* bands in TaC, and C-2*p* and Ti-3*d* bands in TiC compounds. Such overlaps are suggestive of covalent bonding between the electronic states involved. The nearest peak at the negative energy below the Fermi level in TDOS is known as bonding peak, while the nearest peak at the positive energy is the anti-bonding peak. The energy gap between these peaks is called the pseudo-gap which is an indication of electrical stability [50,99–101]. In all the XC (X = Nb, Ta, Ti) compounds bonding and anti-bonding peaks are within 1.5 eV from the Fermi level. The interaction of charges among bonding atoms is very crucial for a material's stability; materials possessing higher number of bonding electrons are structurally more stable [102,103].

The electron-electron interaction parameter of a material, known as the Coulomb pseudopotential, can be estimated from the following relation [104]:

$$\mu^* = \frac{0.26\, N\,(E_F)}{1 + N\,(E_F)} \tag{36}$$

where, $N\,(E_F)$ is the total density of states at the Fermi level of the compound. The electron-electron interaction parameters of NbC, TaC, and TiC are therefore found to be 0.17, 0.16, and 0.15. This value is relatively high for NbC. The repulsive Coulomb pseudopotential is responsible for the reduction of the transition temperature, $T_c$, of superconducting compounds [104–106]. It also measures the degree of electron correlations in a mystem.

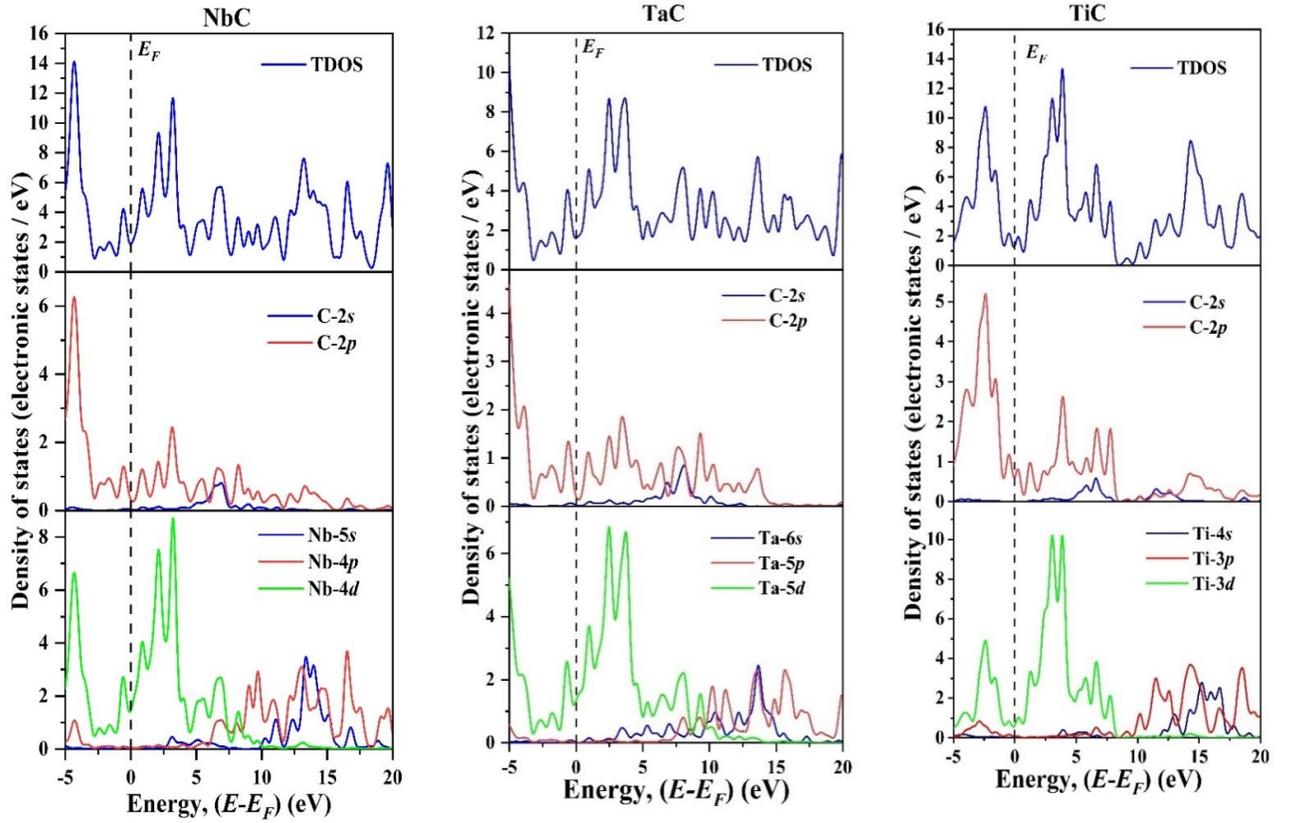

**Fig. 8.** Total and partial density of states of XC (X = Nb, Ta, Ti) compounds.

### *3.4. Fermi surfaces*

In condensed matter physics, the Fermi surface is important to understand the behavior of occupied and unoccupied electronic states of a metallic material at low temperatures. Fermi surface topology of a material dominates a large number of electronic, transport, optical, and magnetic properties. We have constructed the Fermi surfaces of XC (X = Nb, Ta, Ti) compounds from the respective electronic band structures, as shown in Figs.9, 10, and 11. Fermi surfaces are constructed form band number 33, 34, 35, 36, 37, and 38 for NbC, 17, 18, 19, 20, and 21 for TaC and 27, 28, 29, 30, 31, 32, 33, and 34 for TiC, which cross the Fermi level. From Fig. 9, we can observe that electron-like sheet forms around the $\Gamma$-point for band 38 and hole-like sheet forms around the $\Gamma$-point for band 37. From Fig. 10, we can observe that, hole-like sheet forms around the $\Gamma$-point for band 21. From Fig. 11, we can observe that electron-like sheet forms around the $\Gamma$-point for band 30 and hole-like sheet forms around the $\Gamma$-point for band 32. The Fermi surface topologies of NbC and TaC are quite similar. TiC, on the other hand shows distinctly different features. The Fermi sheets are much smaller in this compound compared to the other two.

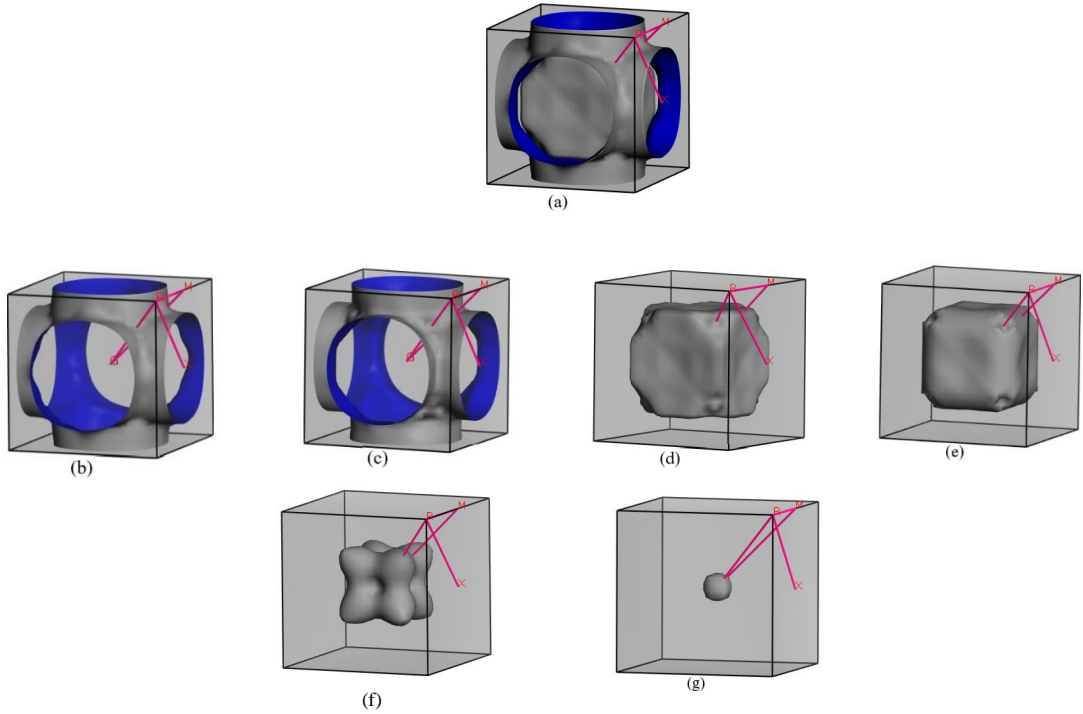

**Fig. 9.** Fermi surface diagram for bands (a) all (b) 33 (c) 34 (d) 35 (e) 36 (f) 37 (g) 38 of NbC.

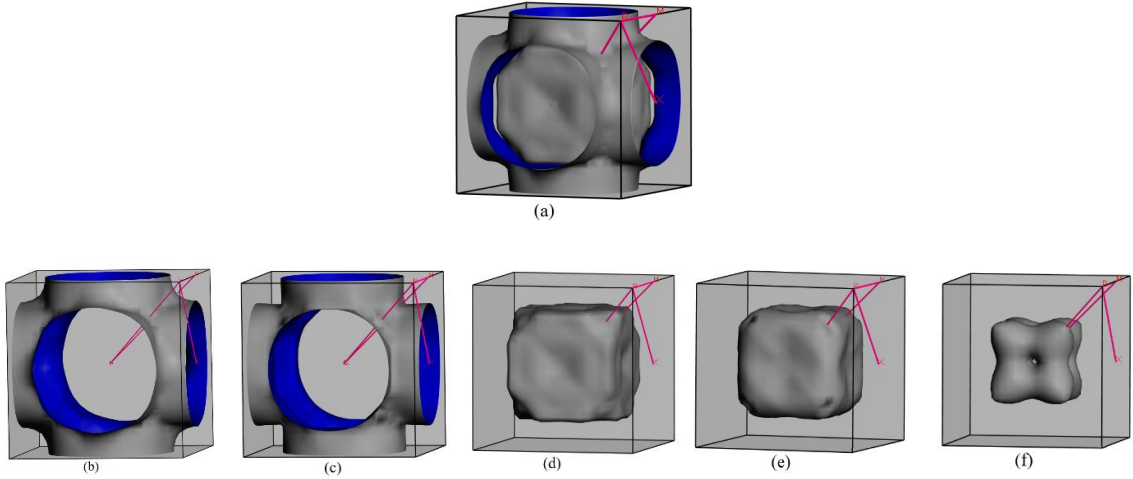

**Fig. 10.** Fermi surface diagram for bands (a) all (b) 17 (c)18 (d) 19 (e) 20 (f) 21 of TaC.

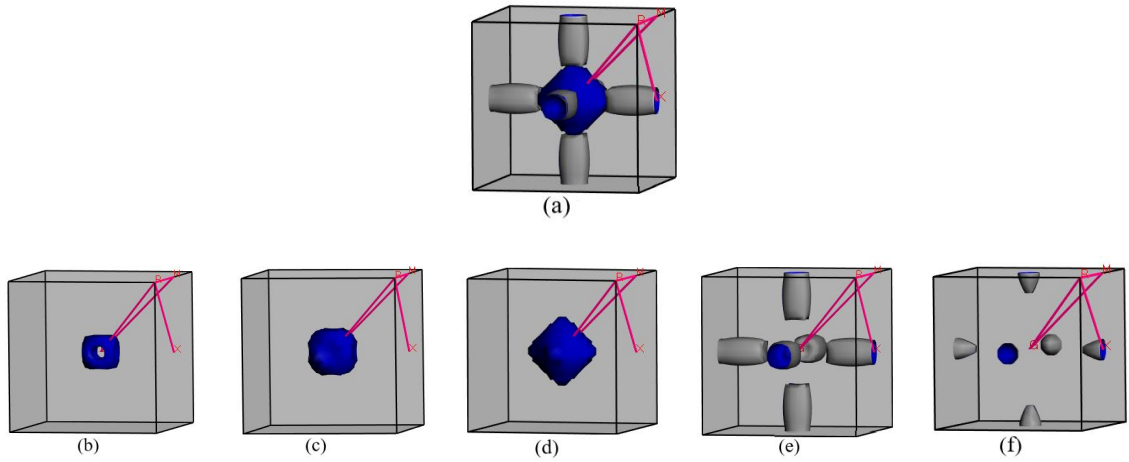

**Fig. 11.** Fermi surface diagram for bands (a) all (b) 30 (c) 31 (d) 32 (e) 33 (f) 34 of TiC.

## *3.5. Charge density*

To study the bonding nature between the atoms of XC (X = Nb, Ta, Ti) directly, the valence electron charge density distributions within the (100) and (111) planes are depicted in Fig. 12. The color scale between the maps represents the total electron density. The charge density distribution map shows that there is an ionic bonding between Nb-C, Ta-C, and Ti-C atoms, and TiC is more ionic than NbC and TaC. Red and blue colors indicate high and low electron density, respectively. So, Ti and Nb atoms have high electron density compared to C atoms for TiC and NbC, respectively. On the other hand, C atoms have high electron density compared to Ta atoms for TaC. From the charge density distribution maps, we see that the charge distribution around the C atomic species gets a significantly spherical shape due to the charge distribution of the surrounding Ti and Nb atoms. This is an indication of ionic bonding. But the charge distribution around the C atomic species gets a significantly distorted shape due to the charge distribution of the surrounding Ta atoms. This is an indication of covalent bonding. Since charge on C atoms is higher than those of the Ta atoms, there is also ionic contribution. So, it appears that TaC possesses ionic and significant covalent bonds.

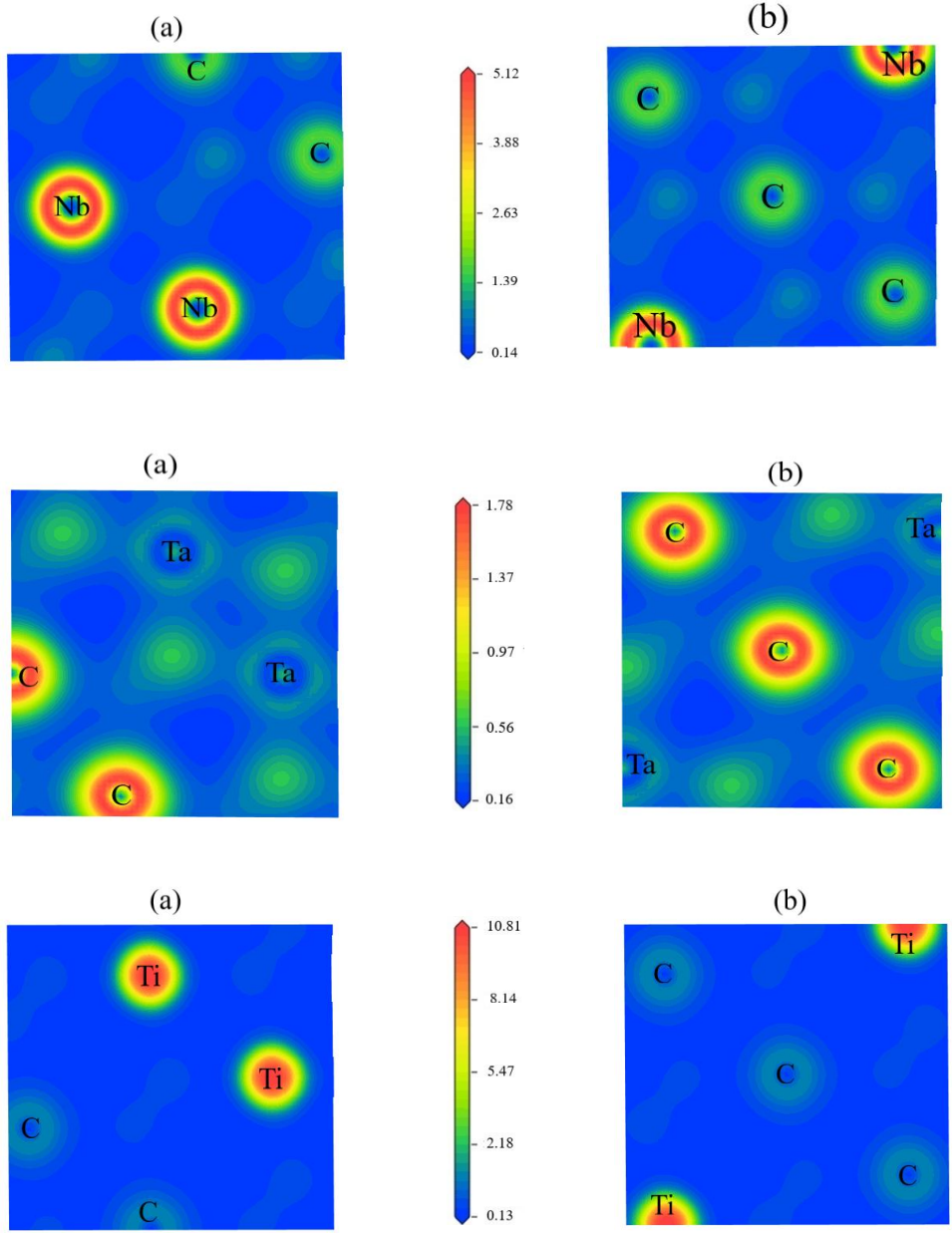

**Fig. 12.** Charge density distribution map of XC (X = Nb, Ta, Ti) in (a) (100) and (b) (111) planes. The charge density scale is given in between panels.

## 3.6. Bond population analysis

To explore the bonding nature (ionic, covalent or metallic) and effective valence of an atom in the molecule in greater depth, we have used both Mulliken population analysis (MPA) [41] and Hirshfeld population analysis (HPA) [107]. The results of this analysis for XC (X = Nb, Ta, Ti) are disclosed in Table 7. The amount of missing valence charges in a projection is represented by the charge spilling parameter. The lower value of charge spilling parameter indicates a good representation of electronic bonds. According to MPA, in XC (X = Nb, Ta, Ti) compounds, Nb, Ta, and Ti atoms transfer 0.71e, 0.74e, and 0.70e charge to C atom respectively. The transfer of electrons between different atoms in the compounds is due to the partial presence of ionic bonding. The difference between formal ionic charge and calculated Mulliken charge is called the effective valence charge (EVC) [43]. Non-zero EVC in all the three compounds is an indication of covalent bonding among the atoms in all these compounds in Table 7. According to HPA, in XC (X = Nb, Ta, Ti) compounds, Nb, Ta, and Ti atoms transfer 0.40e, 0.29e, and 0.27e charge to C atom, respectively. This result is qualitatively consistent with that of MPA. Therefore, MPA and HPA predict that TiC has higher level of covalency compared to NbC and TaC.

**Table 7.** Charge spilling parameter (%), orbital charges (electron), atomic Mulliken charge (electron), Hirshfeld charge (electron), and EVC (electron) of XC (X = Nb, Ta, Ti).

| Compound | Atoms | Charge spilling (%) | s | p | d | Total | Mulliken charge | Hirshfeld charge | EVC |
|---|---|---|---|---|---|---|---|---|---|
| NbC | C | 0.11 | 1.46 | 3.25 | 0.00 | 4.71 | -0.71 | -0.40 | -3.29 |
|  | Nb |  | 2.17 | 6.37 | 3.75 | 12.29 | 0.71 | 0.40 | 4.29 |
| TaC | C | 0.57 | 1.48 | 3.26 | 0.00 | 4.74 | -0.74 | -0.29 | -3.26 |
|  | Ta |  | 0.36 | 0.13 | 3.76 | 4.26 | 0.74 | 0.29 | 4.26 |
| TiC | C | 0.05 | 1.50 | 3.20 | 0.00 | 4.70 | -0.70 | -0.27 | -3.30 |
|  | Ti |  | 2.13 | 6.58 | 2.59 | 11.30 | 0.70 | 0.27 | 2.30 |

## 3.7. Theoretical bond hardness

The hardness of a material always plays an important role in its applications, especially as an abrasive resistant phase and radiation tolerant material [108]. A compound with higher bond density or electronic density, shorter bond length, and greater degree of covalent bonding is harder. In general, a compound with larger bulk modulus and shear modulus indicates higher hardness. Despite the fact that the bulk modulus or shear modulus provide some information on hardness, there is no simple one-to-one relationship between hardness and bulk/shear modulus [109]. There are two types of hardness: intrinsic and extrinsic. In general, the hardness of a perfect crystal is considered as intrinsic, while the hardness of nanocrystalline and polycrystalline aggregate is considered as extrinsic. The hardness of compound can be obtained using the following equations [110,111]:

$$H_v^\mu = \left[\prod^\mu \left\{740(P^\mu - P^{\mu\prime})(v_b^\mu)^{-5/3}\right\}^{n^\mu}\right]^{1/\sum n^\mu} \quad (37)$$

and

$$H_v = \left[\prod^\mu (H_v^\mu)^{n^\mu}\right]^{1/\sum n^\mu} \quad (38)$$

where $P^\mu$ is the Mulliken bond overlap population of the μ-type bond, $P^{\mu\prime} = n_{free}/V$ is the metallic population (with $n_{free} = \int_{E_P}^{E_F} N(E)\, dE$ = the number of free electrons; $E_P$ and $E_F$ are the energy at the pseudogap and at the Fermi level, respectively), $n^\mu$ is the number of μ-type bond, $v_b^\mu$ is the bond volume of μ-type bond calculated by using the equation $v_b^\mu = (d^\mu)^3/\sum_v[(d^\mu)^3 N_b^\mu]$, $H_v^\mu$ is the bond hardness of μ-type bond and $H_v$ is the hardness of the compound. The constant 740 is a proportional coefficient fitted from the hardness of diamond.

The calculated bond length, overlap population and the theoretical hardness of XC (X = Nb, Ta, Ti) are given in Table 8. The hardness of covalent crystal depends on the sum of resistance of each bond per unit area to the indenter [110,112]. The Mulliken bond populations define the degree of overlap of the electron clouds forming bonding between atoms in the crystal. The overlap population of electrons between atoms is measure of the strength of the covalent bond between atoms and the strength of the bond per unit volume. The positive (+) and negative (-) values of overlap population indicate the presence of bonding type and anti-bonding type interactions between the atoms, respectively [113]. The overlap population close to zero indicates that there is no significant interaction between the electronic populations of the two atoms involved. The calculated values indicate that the bonding type interaction is present in XC (X = Nb, Ta, Ti) compounds.

Mulliken population analysis is also useful in determining the metallic nature of bonds in a compound. Metallic bonding is soft in nature and has negligible contribution to the overall hardness of a material [114]. Metallic populations for XC (X = Nb, Ta, Ti) compounds are very low. The metallicity of a crystal can be defined as,

$$f_m = P^{\mu\prime}/P^\mu \quad (39)$$

where, $P^{\mu\prime}$ and $P^\mu$ are metallic population and the Mulliken (overlap) population, respectively. The calculated results are listed in Table 8. The calculated values of hardness $H_v^\mu$ of a μ-type bond and total hardness $H_v$ of XC (X = Nb, Ta, Ti) compounds are all disclosed in Table 8. The obtained hardness is highly consistent with previous literature. As can be seen from Table 8, the hardness of TiC is significantly higher than that of TaC and NbC, because of the smaller bond length and cell volume in TiC.

**Table 8.** The calculated Mulliken bond overlap population of $\mu$-type bond $P^\mu$, bond length $d^\mu$ (Å), metallic population $P^{\mu'}$, metallicity $f_m$, total number of bonds $N^\mu$, cell volume $\Omega$ (Å$^3$), bond volume $v_b^\mu$ (Å$^3$), hardness of $\mu$-type bond $H_v^\mu$ (GPa) and Vickers hardness of the compound $H_v$ (GPa) of XC (X = Nb, Ta, Ti).

| Compound | Bond | $d^\mu$ | $P^\mu$ | $P^{\mu'}$ | $f_m$ | $N^\mu$ | $\Omega$ | $v_b^\mu$ | $H_v^\mu$ | $H_v$ | Ref. |
|---|---|---|---|---|---|---|---|---|---|---|---|
| NbC | C-Nb | 2.24 | 0.76 | 0.02 | 0.03 | 12 | 89.834 | 7.49 | 19.10 | 19.10 | This work |
|  |  | 2.24 | 0.76 | - | - | 12 | 90.189 | 7.52 | 19.50 | 19.50 | [17] |
| TaC | C-Ta | 2.21 | 0.94 | 0.02 | 0.02 | 12 | 86.713 | 7.37 | 24.39 | 24.39 | This work |
|  |  | 2.28 | 0.91 | - | - | 12 | 95.284 | 7.90 | 21.30 | 21.30 | [17] |
| TiC | C-Ti | 2.17 | 0.81 | 0.009 | 0.01 | 12 | 81.204 | 6.77 | 24.47 | 24.47 | This work |
|  |  | 2.17 | 0.82 | - | - | 12 | 81.390 | 6.78 | 25.00 | 25.00 | [17] |

## 3.8. Thermophysical properties

### (a) Debye temperature and sound velocities

The Debye temperature ($\Theta_D$) is one of the most important thermophysical parameters in solids, as it is linked to a wide range of physical properties such as thermal conductivity, lattice vibration, interatomic bonding, vacancy formation energy, melting temperature, coefficient of thermal expansion, and phonon specific heat. Debye temperature can be defined as the temperature at which all the atomic modes of vibrations become activated. Vibrational excitations at low temperatures are solely caused by acoustic modes. As a result, at low temperature $\Theta_D$ calculated from elastic constants and specific heat are identical. This temperature depends on the crystal stiffness and atomic masses of the compound's constituent atoms. Generally, larger Debye temperatures are seen in materials with better interatomic bonding strength, higher melting temperature, greater hardness, higher mechanical wave velocity, and lower average atomic mass. All the modes of vibrations possess an energy $\sim k_B T$ when the temperature is higher than $\Theta_D$. On the other hand, when $T < \Theta_D$, the higher frequency modes are expected to be frozen and quantum nature of vibrational modes are manifested [115]. In this study, $\Theta_D$ is calculated from its proportionality to the mean sound velocity inside the crystal as [116,117]:

$$\Theta_D = \frac{h}{k_B}\left[\left(\frac{3n}{4\pi}\right)\frac{N_A\rho}{M}\right]^{\frac{1}{3}} v_m \qquad (40)$$

where $h$ is the Planck's constant, $k_B$ is the Boltzmann's constant, $n$ denotes number of atoms within the unit cell, $M$ is molar mass, $\rho$ is density, $N_A$ is Avogadro's number and $v_m$ denotes mean sound velocity. $v_m$ can be determined from bulk ($B$) and shear ($G$) modulus through longitudinal ($v_l$) and transverse ($v_t$) sound velocities as follows:

$$v_m = \left[\frac{1}{3}\left(\frac{2}{v_t^3} + \frac{1}{v_l^3}\right)\right]^{-\frac{1}{3}} \tag{41}$$

where,

$$v_t = \sqrt{\frac{G}{\rho}} \tag{42}$$

$$v_l = \sqrt{\frac{3B+4G}{3\rho}} \tag{43}$$

The calculated Debye temperature, $\Theta_D$ and $v_l$, $v_t$, and $v_m$ are enlisted in Table 9. It is well known that a higher Debye temperature implies a higher phonon thermal conductivity. Strong chemical bonding favors higher Debye temperature. As shown in Table 9, $\Theta_D$ value of TiC is higher than that of NbC and TaC. Therefore, we can conclude that the chemical bonds in TiC are stronger than that of NbC and TaC. The estimated values of $\Theta_D$ of XC (X = Nb, Ta, Ti) show very good agreement with previously obtained results.

**Table 9.** Calculated mass density ($\rho$ in gm cm$^{-3}$), longitudinal, transverse and sound velocities ($v_l$, $v_t$ and $v_m$ in ms$^{-1}$), and Debye temperature ($\Theta_D$ in K) of XC (X = Nb, Ta, Ti) compounds.

| Compound | $\rho$ | $v_l$ | $v_t$ | $v_m$ | $\Theta_D$ | Ref. |
|---|---|---|---|---|---|---|
| | 7.76 | 8843.49 | 5431.55 | 5994.50 | 797.83 | This work |
| NbC | 7.73 | 8152.00 | 4615.50 | 5132.70 | 680.9 | [17] |
| | 14.49 | 7059.33 | 4283.59 | 4733.41 | 634.22 | This work |
| TaC | 13.45 | 6189.90 | 3517.40 | 3910.30 | 509.40 | [17] |
| | 4.90 | 10021.78 | 6074.87 | 6713.49 | 924.07 | This work |
| TiC | 4.89 | 10369.50 | 6468.80 | 7145.10 | 980.70 | [17] |

The frequency and dimension of a material have no effect on the velocity of sound waves (longitudinal and transverse), but the nature of the material does. There are only three modes of vibration for each atom in a system (one longitudinal and two transverse modes). In anisotropic crystals, the pure longitudinal and transverse modes are only present along certain crystallographic directions. Sound velocities should be measured in different propagation directions to understand their anisotropic character. Pure transverse and longitudinal modes can be identified for [001], [110], and [111] directions with cubic symmetry; sound propagation modes in other directions are quasi-transverse or quasi-longitudinal. The acoustic velocities in the principle directions for cubic crystal can be expressed as [89]:

$$[100]\upsilon_l = \sqrt{C_{11}/\rho}; \quad [010]\upsilon_{t1} = [001]\upsilon_{t2} = \sqrt{C_{44}/\rho} \quad (44)$$

$$[110]\upsilon_l = \sqrt{(C_{11} + C_{12} + 2C_{44})/2\rho}; \quad [1\bar{1}0]\upsilon_{t1} = \sqrt{C_{11} - C_{12}/\rho}; \quad [001]\upsilon_{t2} = \sqrt{C_{44}/\rho} \quad (45)$$

$$[111]\upsilon_l = \sqrt{(C_{11} + 2C_{12} + 4C_{44})/3\rho}; \quad [11\bar{2}]\upsilon_{t1} = \upsilon_{t2} = \sqrt{C_{11} - C_{12} + C_{44}/3\rho} \quad (46)$$

where $\upsilon_{t1}$ and $\upsilon_{t2}$ refers to the first transverse mode and the second transverse mode, respectively. From these equations we can say that a compound with the value of small density $\rho$ and large elastic constants have large sound velocities. Therefore, TiC have larger sound velocities than that in NbC and TaC. Directional sound velocities of XC (X = Nb, Ta, Ti) are disclosed in Table 10.

**Table 10.** Anisotropic sound velocities (ms$^{-1}$) of XC (X = Nb, Ta, Ti) along different crystallographic directions.

| Propagation directions | | NbC | TaC | TiC |
|---|---|---|---|---|
| [111] | $[111]\upsilon_l$ | 8405.68 | 6627.70 | 9888.12 |
| | $[11\bar{2}]\upsilon_{t1}$ | 5908.52 | 4793.48 | 6149.91 |
| | $[11\bar{2}]\upsilon_{t2}$ | 5908.52 | 4793.48 | 6149.91 |
| [110] | $[110]\upsilon_l$ | 8728.80 | 6961.05 | 9975.43 |
| | $[1\bar{1}0]\upsilon_{t1}$ | 8994.13 | 7417.18 | 8956.70 |
| | $[001]\upsilon_{t2}$ | 4882.37 | 3730.67 | 5908.31 |
| [100] | $[100]\upsilon_l$ | 9633.36 | 7876.90 | 10232.90 |
| | $[010]\upsilon_{t1}$ | 4882.37 | 3730.67 | 5908.31 |
| | $[001]\upsilon_{t2}$ | 4882.37 | 3730.67 | 5908.31 |

**(b) Lattice thermal conductivity**

The lattice thermal conductivity of materials is an important feature to investigate for high-temperature applications. The lattice thermal conductivity ($k_{ph}$) of a solid at different temperatures determines the amount of heat energy carried by lattice vibration. The $k_{ph}$ as a function of temperature can be estimated from a formula derived by Slack [118] given bellow:

$$k_{ph} = A \frac{M_{av} \Theta_D^3 \delta}{\gamma^2 n^{2/3} T} \tag{47}$$

In this formulation, $M_{av}$ is the average atomic mass in kg/mol, $\Theta_D$ is the Debye temperature in K, $\delta$ is the cubic root of average atomic volume in meter (m), $n$ refers to the number of atoms in the conventional unit cell, $T$ is the absolute temperature in K, and $\gamma$ is the acoustic Grüneisen parameter which measures the degree of anharmonicity of phonons. A material with a small value of the Grüneisen parameter indicates low anharmonicity of phonons which results in high thermal conductivity. It is a dimensionless quantity which can be derived from Poisson's ratio using the equation [118]:

$$\gamma = \frac{3(1+v)}{2(2-3v)} \tag{48}$$

The factor $A(\gamma)$, due to Julian [119], is given by:

$$A(\gamma) = \frac{5.720 \times 10^7 \times 0.849}{2 \times (1 - 0.514/\gamma + 0.228/\gamma^2)} \tag{49}$$

The calculated room temperature (300 K) lattice thermal conductivity is listed in Table 11.

**(c) Minimum thermal conductivity**

At high temperatures above the Debye temperature, thermal conductivity of a compound approaches a minimum value known as minimum thermal conductivity ($k_{min}$). The minimal thermal conductivity is notable in that it does not depend on the presence of defects (such as dislocations, individual vacancies and long-range strain fields associated with inclusions and dislocations) inside the crystal. This is due to the fact that these defects affect phonon transport over length scales much larger than the interatomic spacing and at high temperatures, the phonon mean free path becomes significantly smaller than this length scale. Based on the Debye model, Clarke deduced the following formula for calculating the minimum thermal conductivity $k_{min}$ of compounds at high temperatures [120]:

$$k_{min} = k_B \upsilon_m (V_{atomic})^{-2/3} \tag{50}$$

In this equation, $k_B$ is the Boltzmann constant, $\upsilon_m$ is the average sound velocity and $V_{atomic}$ is the cell volume per atom.

The calculated values of minimum thermal conductivity for XC (X = Nb, Ta, Ti) are enlisted in Table 11. $k_{min}$ of TiC is higher than that of NbC and TaC. Compounds with higher sound velocity and Debye temperature have higher minimum thermal conductivity.

Heat is known to be transmitted through solids in three different modes: by thermal vibrations of atoms, by the movement of free electrons in metals and by radiation if they are transparent. The propagation of elastic waves is involved in thermal vibration assisted transmission. An elastically anisotropic material also has anisotropic minimum thermal conductivity. The anisotropy in minimum thermal conductivity depends on sound velocity in different crystallographic directions. The minimum thermal conductivities along different directions are calculated by using Cahill and Clarke model [121]:

$$k_{min} = \frac{k_B}{2.48} n^{2/3} (\upsilon_l + \upsilon_{t1} + \upsilon_{t2}) \tag{51}$$

where, $k_B$ is the Boltzmann constant, $n$ is the number of atoms per unit volume and $N$ is the total number of atoms in the cell having a volume $V$.

The minimum thermal conductivity of XC (X = Nb, Ta, Ti) along [001], [110], and [111] directions are summarized in Table 11. Minimum thermal conductivity of TiC is higher than that of NbC and TaC. For XC (X = Nb, Ta, Ti) compounds, the minimum thermal conductivities along different crystallographic directions are higher than the isotropic minimum thermal conductivity.

**Table 11.** The number of atoms per mole of the compound $n$ (m$^{-3}$), lattice thermal conductivity $k_{ph}$ (W/m.K) at 300 K, minimum thermal conductivity (W/m. K) of XC (X = Nb, Ta, Ti) compounds along different directions.

| Compound | $n$ (10$^{28}$) | $k_{ph}$ | [001] $k_{min}$ | [110] $k_{min}$ | [111] $k_{min}$ | $k_{min}$ |
|---|---|---|---|---|---|---|
| NbC | 8.91 | 93.83 | 2.15 | 2.51 | 2.24 | 1.65 |
| TaC | 9.04 | 85.54 | 1.72 | 2.03 | 1.82 | 1.32 |
| TiC | 9.84 | 80.27 | 2.62 | 2.95 | 2.63 | 1.98 |

**(d) Melting temperature**

The melting temperature is another interesting and significant parameter for solids. In the high temperature applications, it sets the limit. The melting temperature of a crystalline material is related to its bonding energy and thermal expansion. A high value of melting temperature indicates strong atomic bonding, a high value of bonding energy and a low value of thermal expansion. The elastic constants can be used to calculate the melting temperature $T_m$ of solids using the following equation [122]:

$$T_m = 354K + 4.5(K/GPa)\left(\frac{2C_{11}+C_{33}}{3}\right) \pm 300K \tag{53}$$

The calculated melting temperature of XC (X = Nb, Ta, Ti) are disclosed in Table 12. The melting temperature of TaC is higher than that of NbC and TiC, respectively which means TaC is a good candidate material for high temperature application compared to NbC and TiC.

**(e) Thermal expansion coefficient and heat capacity**

A material's thermal expansion coefficient (TEC) is linked to a variety of other physical properties, including specific heat, thermal conductivity, temperature variation of the energy band gap, and electron effective mass. The thermal expansion coefficient ($\alpha$) is an intrinsic thermal property of a material and it is important for epitaxial growth of crystals. The thermal expansion coefficient of a material can be determined using the following equation [123,124]:

$$\alpha = \frac{1.6 \times 10^{-3}}{G} \tag{54}$$

where, $G$ is the shear modulus (in GPa). The thermal expansion coefficient is inversely related to melting temperature: $\alpha \approx 0.02/T_m$ [120,124]. The computed thermal expansion coefficient of XC (X = Nb, Ta, Ti) compounds are disclosed in Table. 12.

Another important intrinsic thermodynamic property of a material is its heat capacity. Thermal conductivity and thermal diffusivity are higher in materials with higher heat capacity. The change in thermal energy per unit volume in a material per Kelvin change in temperature is defined by its heat capacity per unit volume ($\rho C_P$). The heat capacity of a material per unit volume can be calculated from [123]:

$$\rho C_P = \frac{3k_B}{\Omega} \tag{55}$$

The heat capacity per unit volume of XC (X = Nb, Ta, Ti) materials are shown in Table 12. The heat capacity of TiC is higher than that of TaC and NbC. So, TiC is expected to possess comparatively higher thermal conductivity.

**(f) Dominant phonon mode**

Phonons play an important role in determining variety of physical properties of condensed matter, such as heat capacity, thermal conductivity and electrical conductivity. The wavelength of the dominant phonon is the wavelength, $\lambda_{dom}$, at which the phonon distribution attains a peak. The wavelength of the dominant phonon for XC (X = Nb, Ta, Ti) at 300 K has been calculated by the following relationship [52,120]:

$$\lambda_{dom} = \frac{12.566\, v_m}{T} \times 10^{-12} \tag{56}$$

where, $v_m$ is the average sound velocity in ms$^{-1}$, $T$ is the temperature in K. A material with higher average sound velocity, higher shear modulus, lower density has higher wavelength of the dominant phonon. The estimated values of $\lambda_{dom}$ in meter are listed in Table 12.

**Table 12.** Calculated melting temperature $T_m$ (K), thermal expansion coefficient α ($K^{-1}$), heat capacity per unit volume $\rho C_p$ (JK$^{-1}$m$^{-3}$) and wavelength of the dominant phonon mode at 300 K $\lambda_{dom}$ (m), of XC (X = Nb, Ta, Ti) compounds.

| Compound | $T_m$ | α ($10^{-6}$) | $\rho C_p$ ($10^6$) | $\lambda_{dom}$ ($10^{-12}$) |
|---|---|---|---|---|
| NbC | 3594.63 | 6.99 | 3.69 | 251.09 |
| TaC | 4399.68 | 6.02 | 3.82 | 198.27 |
| TiC | 2662.91 | 8.85 | 4.08 | 281.21 |

## *3.9. Optical properties*

The understanding of energy/frequency dependent optical parameters is essential to predict how a material will respond when electromagnetic radiation is incident on it. In recent decades, interest on optical properties of materials has increased many folds in materials science because of their close relations to the applications in integrated optics such as optical modulation, optoelectronics, optical information processing and communications, display devices, and optical data storage. The energy or frequency dependent optical properties also provide with an important tool for studying energy band structure, state of impurity levels, excitons, localized defects, lattice vibrations, and certain magnetic excitations [125,126]. In order to investigate possible optoelectronic applications of a compound, knowledge regarding the response of the compound to infrared, visible and ultraviolet spectra is important.

Various frequency dependent optical constants, namely, dielectric function $\varepsilon(\omega)$, refractive index $n(\omega)$, optical conductivity $\sigma(\omega)$, reflectivity $R(\omega)$, absorption coefficient $\alpha(\omega)$ and energy loss function $L(\omega)$ (where $\omega = 2\pi f$ is the angular frequency) are calculated in this section to explore the response of XC (X = Nb, Ta, Ti) to incident photons. The optical parameters of the compounds are depicted in Figs.13, 14, and 15 for incident energies up to 30 eV and the electric field polarizations along the [100] directions.

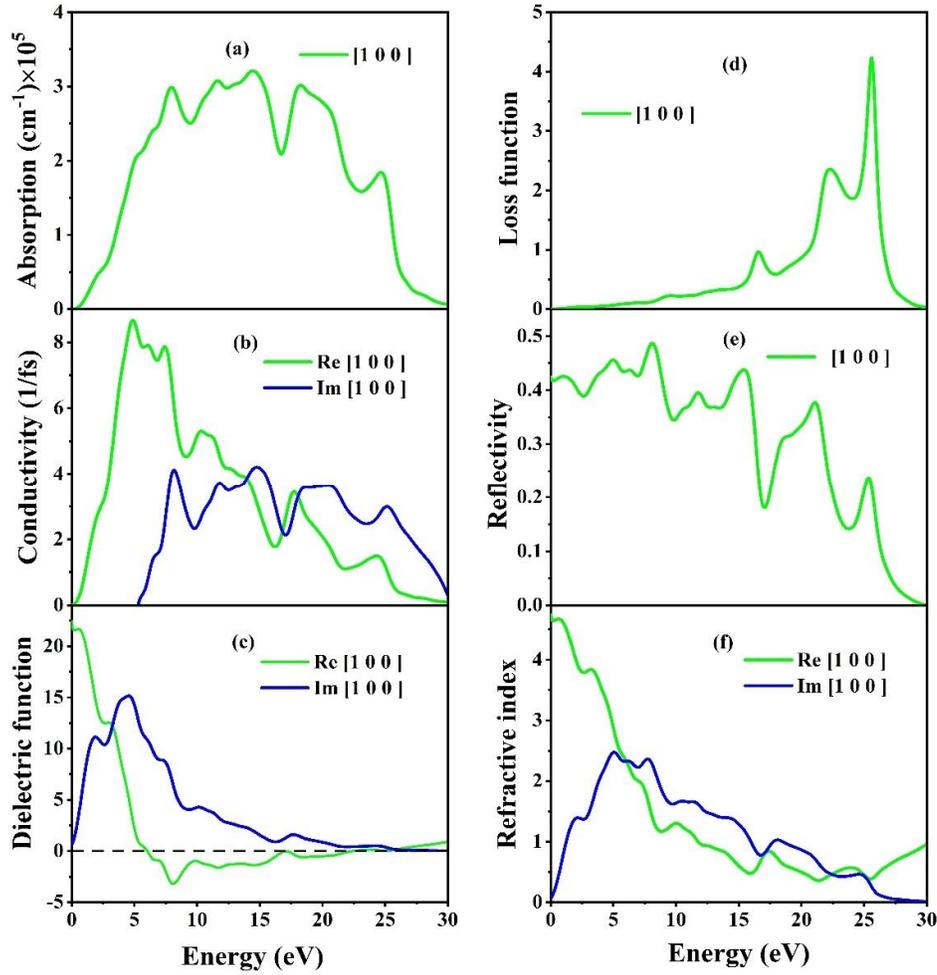

**Figure 13.** The frequency dependent (a) absorption coefficient (b) optical conductivity (c) dielectric function (d) loss function (e) reflectivity, and (f) refractive index of NbC with electric field polarization vectors along [100] direction.

The absorption coefficient is an important parameter to understand the optimum solar energy conversion efficiency of a material. It also reveals the electrical nature of the material, whether it is metallic, semiconducting or insulating. Figs. 13(a), 14(a), and 15(a) show the absorption coefficient α(ω) of XC (X = Nb, Ta, Ti) compounds. The absorption coefficients of XC (X = Nb, Ta, Ti) start from 0 eV, which is an indication of its metallic band structure. The absorption coefficient of NbC is quite high in the spectral region from ~ 5.4 to 24.8 eV, peaking around 14.5 eV. The absorption coefficient of TaC is high in the spectral region from ~ 5.96 to 29 eV, peaking around 12.7 eV. The absorption coefficient of TiC is also quite high in the spectral region from ~ 1.52 to 25 eV, peaking around 19.1 eV. α(ω) decreases sharply at ~ 25.2 eV, ~ 29.2 eV, and ~ 25.6 eV for NbC, TaC, and TiC, respectively, in agreement with the position of the loss peak.

The conduction of free charge carriers over a defined range of the photon energies are explained by the optical conductivity. This is a dynamic response of mobile charge carriers which includes the photon generated electron hole pairs in semiconductors. Figs. 13(b), 14(b), and 15(b) show the real and imaginary parts of optical conductivity $\sigma(\omega)$. For XC (X = Nb, Ta, Ti), optical conductivity starts with zero photon energy, which indicates that the materials have no band gap agreeing with the band structure and TDOS calculations and indicating once again its metallic character.

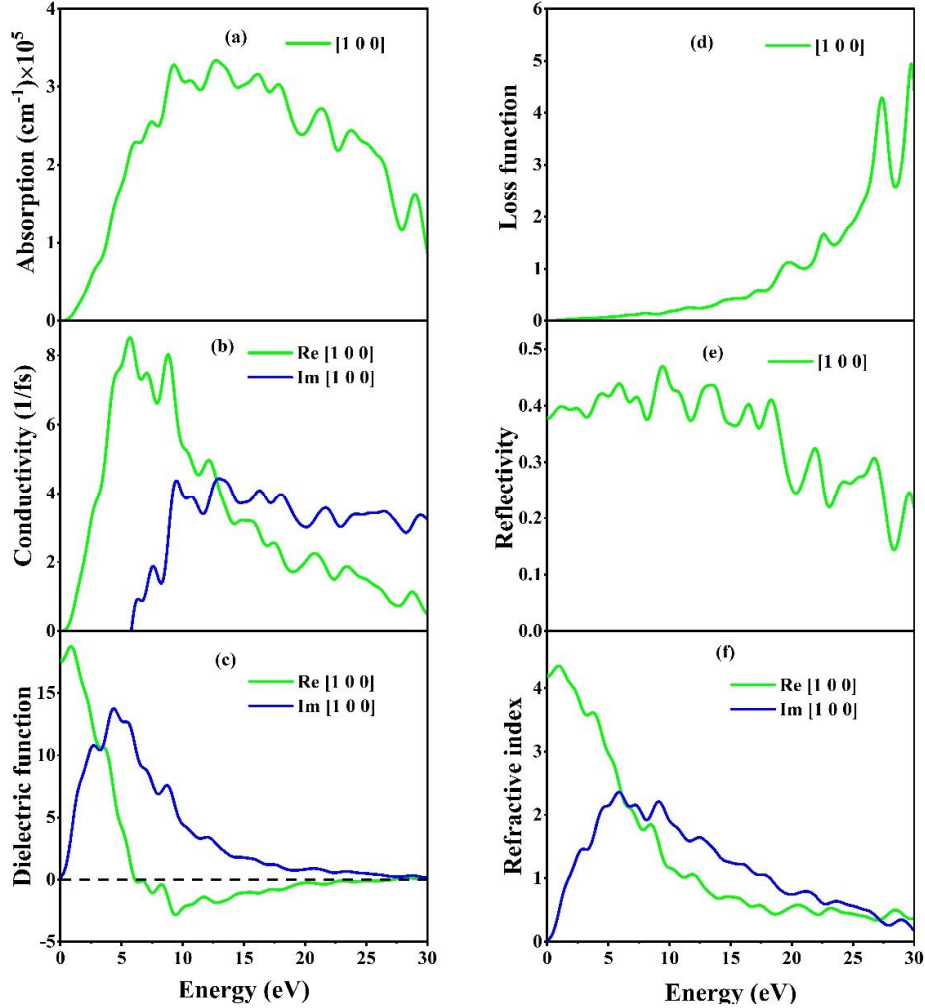

**Figure 14.** The frequency dependent (a) absorption coefficient (b) optical conductivity (c) dielectric function (d) loss function (e) reflectivity, and (f) refractive index of TaC with electric field polarization vectors along [100] direction.

The real and imaginary parts of dielectric constants are illustrated in Figs. 13(c), 14(c), and 15(c). The real part of the dielectric constant is related to the electrical polarization of the material, while the imaginary part is linked with dielectric loss. The real part crosses zero from below at ~22.4 eV for NbC and at ~27.2 for TaC and at ~22.4 eV for TiC. After that this part gradually approaches unity.

The loss functions $L(\omega)$ of XC (X = Nb, Ta, Ti) are depicted in Figs. 13(d), 14(d), and 15(d). The loss peaks are found at ~25.2 eV for NbC and at ~29.2 for TaC and at ~25.6 eV for TiC.

These peaks mark the characteristic plasmon energy for corresponding materials. The plasma oscillations due to collective motions of the charge carriers are induced at these particular energies. It is worth noticing that the plasma energies coincide with the sharp falls in the absorption coefficient and reflectivity. This implies that the compounds under investigation are expected to behave transparently for photons with energies greater than the plasma energy and the optical properties will show behaviors similar to those for insulating systems.

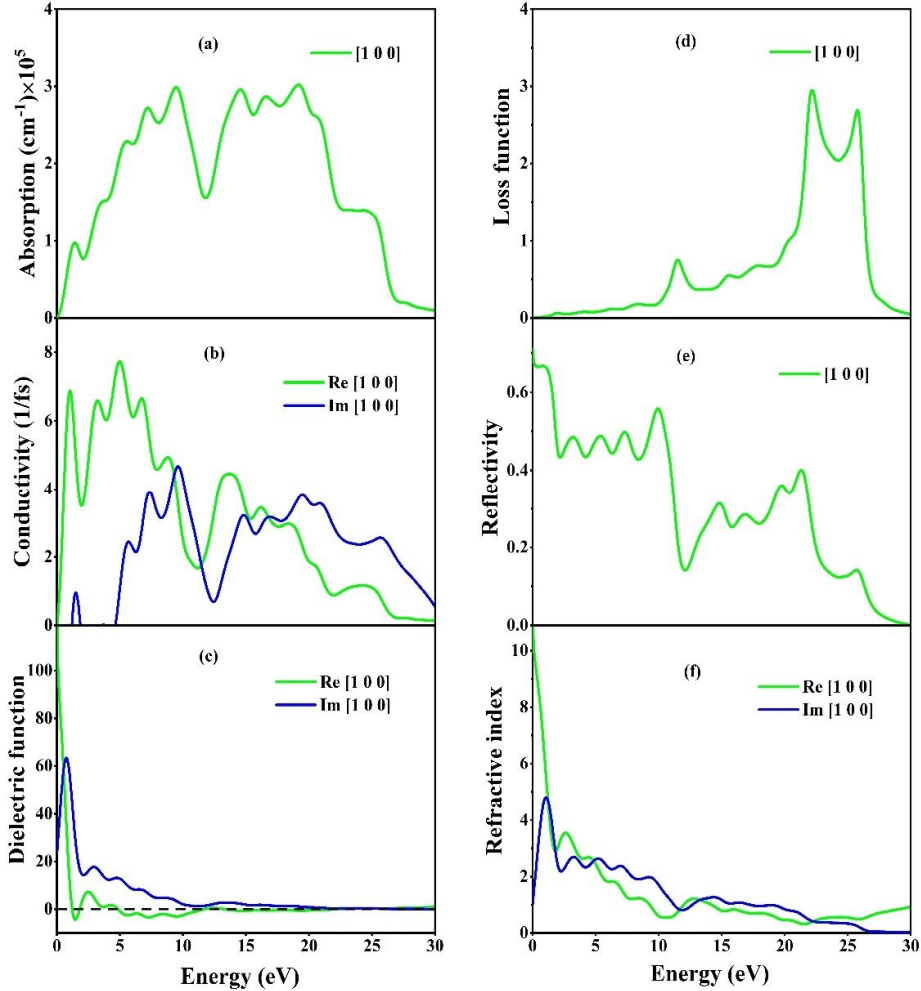

**Figure 15.** The frequency dependent (a) absorption coefficient (b) optical conductivity (c) dielectric function (d) loss function (e) reflectivity, and (f) refractive index of TiC with electric field polarization vectors along [100] direction.

The reflectivity profile for XC (X = Nb, Ta, Ti) compounds are shown in Figs. 13(e), 14(e), and 15(e). The reflectivity spectra fall sharply at plasma frequency in all three compounds. In NbC, the maximum reflectivity value is about 49% at ~8 eV. The reflectivity spectra stay above 40% from infrared to near-ultraviolet region. $R(\omega)$ falls sharply at ~25.2 eV as the material becomes transparent to the incident EMW. In TaC, $R(\omega)$ remains over 40% in the energy range from 0 eV to ~18.5 eV. So, this material has a wide band reflectivity and can be employed as a reflector to reduce solar heating. TiC is also a good reflector of infrared and near visible light. The reflectivity falls below 45% and remains almost constant (with little oscillations) at near ultraviolet region and rises to a peak around 10 eV. $R(\omega)$ falls sharply at ~25.6 eV as the material becomes transparent to the incident EMW.

The frequency dependent real and imaginary parts of refractive index are represented in Figs. 13(f), 14(f), and 15(f). The phase velocity of electromagnetic wave in the material is determined by the real part of refractive index, whereas the attenuation of electromagnetic radiation inside the material is measured by the imaginary part which is often referred to as extinction coefficient. It is instructive to note that NbC and TaC have high value of the real part of refractive index at low energies covering infrared to visible regions whereas imaginary part of the refractive index remains zero. The real part of refractive index of TiC has very high value at low energies covering the infrared regions compared to NbC and TaC. High refractive index materials have desired optical characteristics for photonic crystals and optoelectronic display devices. The real part of refractive index has high value at low energies (0-4.0 eV) for both NbC and TaC and at low energies (0-1.20 eV) for TiC.

The optical parameters showed herein are novel results and exhibit optical isotropies with respect to the polarization direction of the incident EMW.

## 4. Conclusions

We have investigated the structural, mechanical, electronic, optical, and some thermophysical properties of technologically important XC (X = Nb, Ta, Ti) binary metallic carbides extensively employing the density functional theory based first-principles method. Most of the results regarding elastic, bonding, thermophysical and optical properties are novel. We have also compared our results with earlier theoretical and experimental values where available. A good accord has been found.

All three compounds are mechanically and dynamically stable with brittle characteristics. The compounds are hard. TaC in particular with its high value of hardness, high level of machinability, and extremely high melting temperature can be used as heavy duty engineering tool suitable to be used in very harsh conditions. The thermo-mechanical properties of TaC is much better than those of the industrially important MAX and MAB phase nanolaminates [127-129]. The chemical bondings present in XC (X = Nb, Ta, Ti) are mixed in nature – ionic and covalent contributions dominate. The values of Poisson's ratio ($\sigma$) indicate that non-central forces are present in the materials under study. Based on the values of different elastic anisotropy factors ($A, A_1, A_2, A^U, A^{eq}$, etc.), we can predict that these materials possess little elastic anisotropy. The electronic band structure and total density of states analyses suggest that XC (X = Nb, Ta, Ti) compounds are metal with moderate TDOS at the Fermi level. The calculated Debye temperatures are high for all the XC (X = Nb, Ta, Ti) compounds indicating their hard nature. High values of estimated melting temperatures and minimum phonon thermal conductivity correspond very well to the calculated values of hardness and various other calculated elastic moduli. Clear correspondence among Debye temperature ($\Theta_D$), lattice thermal conductivity ($K_{ph}$), melting temperature ($T_m$), and minimum thermal conductivity ($k_{min}$) are found. The optical parameters show metallic behavior in agreement with the electronic band structure. The real part of the refractive index of TiC has a very high value at low energies covering infrared to visible regions compared to that of NbC and TaC. The real part of the refractive index has a high value at low energies (0-6.0 eV) for both NbC and TaC and at low energies (0-1.20 eV) for TiC. All the compounds under study are efficient absorber of ultraviolet photons.

## Acknowledgments

S. H. N. and R. S. I. acknowledge the research grant (1151/5/52/RU/Science-07/19-20) from the Faculty of Science, University of Rajshahi, Bangladesh, which partly supported this work. S. H. N. dedicates this work in the loving memory of his father, Professor A. K. M. Mohiuddin, who has passed away recently.

## Declaration of interest

The authors declare that they have no known competing financial interests or personal relationships that could have appeared to influence the work reported in this paper.

## Data availability

The data sets generated and/or analyzed in this study are available from the corresponding author on reasonable request.

## References

[1] A. Friedrich, B. Winkler, E. A. Juarez-Arellano, and L. Bayarjargal, *Synthesis of Binary Transition Metal Nitrides, Carbides and Borides from the Elements in the Laser-Heated Diamond Anvil Cell and Their Structure-Property Relations*, Materials **4**, 1648 (2011).

[2] Y.-M. Kim and B.-J. Lee, *Modified Embedded-Atom Method Interatomic Potentials for the Ti–C and Ti–N Binary Systems*, Acta Materialia **56**, 3481 (2008).

[3] L. E. Toth, *Transition Metal Carbides and Nitrides* (Academic Press, New York, 1971).

[4] D. J. Singh and B. M. Klein, *Electronic Structure, Lattice Stability, and Superconductivity of CrC*, Phys. Rev. B **46**, 14969 (1992).

[5] T. Matsuda and H. Matsubara, *Thermophysical and Elastic Properties of Titanium Carbonitrides Containing Molybdenum and Tungsten*, Journal of Alloys and Compounds **562**, 90 (2013).

[6] N. A. Dubrovinskaia, L. S. Dubrovinsky, S. K. Saxena, M. Selleby, and B. Sundman, *Thermal Expansion and Compressibility of $Co_6W_6C$*, Journal of Alloys and Compounds **285**, 242 (1999).

[7] C. Jiang, *First-Principles Study of Structural, Elastic, and Electronic Properties of Chromium Carbides*, Appl. Phys. Lett. **92**, 041909 (2008).

[8] K. Schwarz, *Band Structure and Chemical Bonding in Transition Metal Carbides and Nitrides*, Critical Reviews in Solid State and Materials Sciences **13**, 211 (1987).

[9] W. S. Williams, *Cubic Carbides: Solid-State Physics Explores Materials Having Ionic Structure, Metallic Conductivity, and Covalent Hardness.*, Science **152**, 34 (1966).

[10] W. S. Williams, *Transition-Metal Carbides*, Progress in Solid State Chemistry **6**, 57 (1971).

[11] W. S. Williams, *Physics of Transition Metal Carbides*, Materials Science and Engineering: A **105–106**, 1 (1988).

[12] H. O. Pierson, *Handbook of Refractory Carbides and Nitrides*, 1st ed. (Park Ridge, UT, USA, 1996).

[13] S. T. Oyama, *Introduction to the Chemistry of Transition Metal Carbides and Nitrides* (Springer Netherlands, Dordrecht, 1996).

[14] M. Råsander and A. Delin, *Carbon Vacancy Formation in Binary Transition Metal Carbides from Density Functional Theory*, (2018).

[15] M. Råsander, E. Lewin, O. Wilhelmsson, B. Sanyal, M. Klintenberg, O. Eriksson, and U. Jansson, *Carbon Release by Selective Alloying of Transition Metal Carbides*, Journal of Physics. Condensed Matter : An Institute of Physics Journal **23**, 355401 (2011).

[16] H. H. Hwu and J. G. Chen, *Surface Chemistry of Transition Metal Carbides*, Chem. Rev. **105**, 185 (2005).

[17] Y. Liu, Y. Jiang, R. Zhou, and J. Feng, *First Principles Study the Stability and Mechanical Properties of MC (M=Ti, V, Zr, Nb, Hf and Ta) Compounds*, Journal of Alloys and Compounds **582**, 500 (2014).

[18] T. Amriou, B. Bouhafs, H. Aourag, B. Khelifa, S. Bresson, and C. Mathieu, *FP-LAPW Investigations of Electronic Structure and Bonding Mechanism of NbC and NbN Compounds*, Physica B: Condensed Matter **325**, 46 (2003).

[19] L. López-de-la-Torre, B. Winkler, J. Schreuer, K. Knorr, and M. Avalos-Borja, *Elastic Properties of Tantalum Carbide (TaC)*, Solid State Communications **134**, 245 (2005).

[20] J. E. Lowther, *Binary and Ternary Metal Carbides Based upon Ti, Zr and Hf*, Journal of Physics and Chemistry of Solids **66**, 1064 (2005).

[21] E. G. Maksimov, S. V. Ebert, M. V. Magnitskaya, and A. E. Karakozov, *Ab Initio Calculations of the Physical Properties of Transition Metal Carbides and Nitrides and Possible Routes to High-Tc Superconductivity*, J. Exp. Theor. Phys. **105**, 642 (2007).

[22] A. Sevy, D. J. Matthew, and M. D. Morse, *Bond Dissociation Energies of TiC, ZrC, HfC, ThC, NbC, and TaC*, The Journal of Chemical Physics **149**, 044306 (2018).

[23] A. Srivastava, M. Chauhan, and R. K. Singh, *High-Pressure Phase Transitions in Transition Metal Carbides XC (X = Ti, Zr, Hf, V, Nb, Ta): A First-Principle Study*, Phase Transitions **84**, 58 (2011).

[24] T. Shang, J. Z. Zhao, D. J. Gawryluk, M. Shi, M. Medarde, E. Pomjakushina, and T. Shiroka, *Superconductivity and Topological Aspects of the Rocksalt Carbides NbC and TaC*, Phys. Rev. B **101**, 214518 (2020).

[25] A. Teresiak and H. Kubsch, *X-Ray Investigations of High Energy Ball Milled Transition Metal Carbides*, Nanostructured Materials **6**, 671 (1995).

[26] D. Y. Dang, J. L. Fan, and H. R. Gong, *Thermodynamic and Mechanical Properties of TiC from* Ab Initio *Calculation*, Journal of Applied Physics **116**, 033509 (2014).

[27] P. Yang, H. Fu, X. Guo, B. Rachid, and J. Lin, *Mechanism of NbC as Heterogeneous Nucleus of M₃C in CADI: First Principle Calculation and Experiment Research*, Journal of Materials Research and Technology **9**, 3109 (2020).

[28] W. Kohn and L. J. Sham, *Self-Consistent Equations Including Exchange and Correlation Effects*, Phys. Rev. **140**, A1133 (1965).

[29] S. J. Clark, M. D. Segall, C. J. Pickard, P. J. Hasnip, M. I. J. Probert, K. Refson, and M. C. Payne, *First Principles Methods Using CASTEP*, Zeitschrift Für Kristallographie - Crystalline Materials **220**, 567 (2005).

[30] M. Lewin, E. H. Lieb, and R. Seiringer, *The Local Density Approximation in Density Functional Theory*, Pure Appl. Analysis **2**, 35 (2020).

[31] V. Sahni, K.-P. Bohnen, and M. K. Harbola, *Analysis of the Local-Density Approximation of Density-Functional Theory*, Phys. Rev. A **37**, 1895 (1988).

[32] J. P. Perdew, K. Burke, and M. Ernzerhof, *Generalized Gradient Approximation Made Simple*, Phys. Rev. Lett. **77**, 3865 (1996).

[33] K. Burke, J. P. Perdew, and Y. Wang, *Derivation of a Generalized Gradient Approximation: The PW91 Density Functional*, in *Electronic Density Functional Theory: Recent Progress and New Directions*, edited by J. F. Dobson, G. Vignale, and M. P. Das (Springer US, Boston, MA, 1998), pp. 81–111.

[34] J. P. Perdew, J. A. Chevary, S. H. Vosko, K. A. Jackson, M. R. Pederson, D. J. Singh, and C. Fiolhais, *Atoms, Molecules, Solids, and Surfaces: Applications of the Generalized Gradient Approximation for Exchange and Correlation*, Phys. Rev. B **46**, 6671 (1992).

[35] D. Vanderbilt, *Soft Self-Consistent Pseudopotentials in a Generalized Eigenvalue Formalism*, Phys. Rev. B **41**, 7892 (1990).

[36] T. H. Fischer and J. Almlof, *General Methods for Geometry and Wave Function Optimization*, J. Phys. Chem. **96**, 9768 (1992).

[37] O. H. Nielsen and R. M. Martin, *First-Principles Calculation of Stress*, Phys. Rev. Lett. **50**, 697 (1983).

[38] J. P. Watt, *Hashin-Shtrikman Bounds on the Effective Elastic Moduli of Polycrystals with Orthorhombic Symmetry*, Journal of Applied Physics **50**, 6290 (1979).

[39] J. P. Watt and L. Peselnick, *Clarification of the Hashin-Shtrikman Bounds on the Effective Elastic Moduli of Polycrystals with Hexagonal, Trigonal, and Tetragonal Symmetries*, Journal of Applied Physics **51**, 1525 (1980).

[40] S. Saha, T. P. Sinha, and A. Mookerjee, *Electronic Structure, Chemical Bonding, and Optical Properties of Paraelectric*, Phys. Rev. B **62**, 8828 (2000).

[41] R. S. Mulliken, *Electronic Population Analysis on LCAO–MO Molecular Wave Functions. I*, J. Chem. Phys. **23**, 1833 (1955).

[42] D. Sanchez-Portal, E. Artacho, and J. M. Soler, *Projection of Plane-Wave Calculations into Atomic Orbitals*, Solid State Communications **95**, 685 (1995).

[43] M. D. Segall, R. Shah, C. J. Pickard, and M. C. Payne, *Population Analysis of Plane-Wave Electronic Structure Calculations of Bulk Materials*, Phys. Rev. B **54**, 16317 (1996).

[44] M. Mattesini, R. Ahuja, and B. Johansson, *Cubic $Hf_3N_4$ and $Zr_3N_4$ : A Class of Hard Materials*, Phys. Rev. B **68**, 184108 (2003).

[45] R. Hill, *The Elastic Behaviour of a Crystalline Aggregate*, Proc. Phys. Soc. A **65**, 349 (1952).

[46] A. Gueddouh, B. Bentria, and I. K. Lefkaier, *First-Principle Investigations of Structure, Elastic and Bond Hardness of $Fe_xB$ (X=1, 2, 3) under Pressure*, Journal of Magnetism and Magnetic Materials **406**, 192 (2016).
[47] M. Jamal, S. Jalali Asadabadi, I. Ahmad, and H. A. Rahnamaye Aliabad, *Elastic Constants of Cubic Crystals*, Computational Materials Science **95**, 592 (2014).
[48] A. Sari, G. Merad, and H. Si Abdelkader, *Ab Initio Calculations of Structural, Elastic and Thermal Properties of $TiCr_2$ and $(Ti,Mg)(Mg,Cr)_2$ Laves Phases*, Computational Materials Science **96**, 348 (2015).
[49] M. I. Naher, F. Parvin, A. K. M. A. Islam, and S. H. Naqib, *Physical Properties of Niobium-Based Intermetallics ($Nb_3B$; B = Os, Pt, Au): A DFT-Based Ab-Initio Study*, Eur. Phys. J. B **91**, 289 (2018).
[50] M. I. Naher and S. H. Naqib, *Structural, Elastic, Electronic, Bonding, and Optical Properties of Topological $CaSn_3$ Semimetal*, Journal of Alloys and Compounds **829**, 154509 (2020).
[51] M. Rajagopalan, S. Praveen Kumar, and R. Anuthama, *FP-LAPW Study of the Elastic Properties of $Al_2X$ (X = Sc,Y,La,Lu)*, Physica B: Condensed Matter **405**, 1817 (2010).
[52] M. I. Naher, M. A. Afzal, and S. H. Naqib, *A Comprehensive DFT Based Insights into the Physical Properties of Tetragonal Superconducting $Mo_5PB_2$*, Results in Physics **28**, 104612 (2021).
[53] W. Kim, *Strategies for Engineering Phonon Transport in Thermoelectrics*, Journal of Materials Chemistry C **3**, 10336 (2015).
[54] S. F. Pugh, *XCII. Relations between the Elastic Moduli and the Plastic Properties of Polycrystalline Pure Metals*, The London, Edinburgh, and Dublin Philosophical Magazine and Journal of Science **45**, 823 (1954).
[55] V. V. Bannikov, I. R. Shein, and A. L. Ivanovskii, *Elastic Properties of Antiperovskite-Type Ni-Rich Nitrides $MNNi_3$ (M=Zn, Cd, Mg, Al, Ga, In, Sn, Sb, Pd, Cu, Ag and Pt) as Predicted from First-Principles Calculations*, Physica B: Condensed Matter **405**, 4615 (2010).
[56] Z. Yang, D. Shi, B. Wen, R. Melnik, S. Yao, and T. Li, *First-Principle Studies of Ca–X (X = Si,Ge,Sn,Pb) Intermetallic Compounds*, Journal of Solid State Chemistry **183**, 136 (2010).
[57] Md. Mahamudujjaman, Md. A. Afzal, R. S. Islam, and S. H. Naqib, *First-Principles Insights into Mechanical, Optoelectronic, and Thermo-Physical Properties of Transition Metal Dichalcogenides $ZrX_2$ (X = S, Se, and Te)*, AIP Advances **12**, 025011 (2022).
[58] G. Vaitheeswaran, V. Kanchana, A. Svane, and A. Delin, *Elastic Properties of $MgCNi_3$ – a Superconducting Perovskite*, J. Phys.: Condens. Matter **19**, 326214 (2007).
[59] M. A. Hadi, M. Zahanggir Alam, I. Ahmed, A. M. M. Tanveer Karim, S. H. Naqib, A. Chroneos, and A. K. M. A. Islam, *A Density Functional Theory Approach to the Effects of C and N Substitution at the B-Site of the First Boride MAX Phase $Nb_2SB$*, Materials Today Communications **29**, 102910 (2021).
[60] G. N. Greaves, A. L. Greer, R. S. Lakes, and T. Rouxel, *Poisson's Ratio and Modern Materials*, Nature Mater **10**, 823 (2011).

[61] O. L. Anderson and H. H. Demarest, *Elastic Constants of the Central Force Model for Cubic Structures: Polycrystalline Aggregates and Instabilities*, J. Geophys. Res. **76**, 1349 (1971).

[62] M. A. Hadi, S. H. Naqib, S.-R. G. Christopoulos, A. Chroneos, and A. K. M. A. Islam, *Mechanical Behavior, Bonding Nature and Defect Processes of $Mo_2ScAlC_2$: A New Ordered MAX Phase*, Journal of Alloys and Compounds **724**, 1167 (2017).

[63] W. Feng and S. Cui, *Mechanical and Electronic Properties of $Ti_2AlN$ and $Ti_4AlN_3$: A First-Principles Study*, Can. J. Phys. **92**, 1652 (2014).

[64] D. G. Pettifor, *Theoretical Predictions of Structure and Related Properties of Intermetallics*, Materials Science and Technology **8**, 345 (1992).

[65] Z. Sun, D. Music, R. Ahuja, and J. M. Schneider, *Theoretical Investigation of the Bonding and Elastic Properties of Nanolayered Ternary Nitrides*, Phys. Rev. B **71**, 193402 (2005).

[66] A. Bouhemadou, *First-Principles Study of Structural, Electronic and Elastic Properties of $Nb_4AlC_3$*, Braz. J. Phys. **40**, 52 (2010).

[67] N. Miao, B. Sa, J. Zhou, and Z. Sun, *Theoretical Investigation on the Transition-Metal Borides with $Ta_3B_4$-Type Structure: A Class of Hard and Refractory Materials*, Computational Materials Science **50**, 1559 (2011).

[68] X.-Q. Chen, H. Niu, D. Li, and Y. Li, *Modeling Hardness of Polycrystalline Materials and Bulk Metallic Glasses*, Intermetallics **19**, 1275 (2011).

[69] Y. Tian, B. Xu, and Z. Zhao, *Microscopic Theory of Hardness and Design of Novel Superhard Crystals*, International Journal of Refractory Metals and Hard Materials **33**, 93 (2012).

[70] D. M. Teter, *Computational Alchemy: The Search for New Superhard Materials*, MRS Bull. **23**, 22 (1998).

[71] E. Mazhnik and A. R. Oganov, *A Model of Hardness and Fracture Toughness of Solids*, Journal of Applied Physics **126**, 125109 (2019).

[72] L. Vitos, P. A. Korzhavyi, and B. Johansson, *Stainless Steel Optimization from Quantum Mechanical Calculations*, Nature Mater **2**, 1 (2003).

[73] R. C. Lincoln, K. M. Koliwad, and P. B. Ghate, *Morse-Potential Evaluation of Second- and Third-Order Elastic Constants of Some Cubic Metals*, Phys. Rev. **157**, 463 (1967).

[74] M. J. Phasha, P. E. Ngoepe, H. R. Chauke, D. G. Pettifor, and D. Nguyen-Mann, *Link between Structural and Mechanical Stability of Fcc- and Bcc-Based Ordered Mg–Li Alloys*, Intermetallics **18**, 2083 (2010).

[75] K. J. Puttlitz and K. A. Stalter, in *Handbook of Lead-Free Solder Technology for Microelectronic Assemblies* (Springer, New York, 2005), p. 95.

[76] L. Kleinman, *Deformation Potentials in Silicon. I. Uniaxial Strain*, Phys. Rev. **128**, 2614 (1962).

[77] S.-H. Jhi, J. Ihm, S. G. Louie, and M. L. Cohen, *Electronic Mechanism of Hardness Enhancement in Transition-Metal Carbonitrides*, Nature **399**, 6732 (1999).

[78] M. M. Hossain, M. A. Ali, M. M. Uddin, A. K. M. A. Islam, and S. H. Naqib, *Origin of High Hardness and Optoelectronic and Thermo-Physical Properties of Boron-Rich Compounds $B_6X$ (X = S, Se): A Comprehensive Study via DFT Approach*, Journal of Applied Physics **129**, 175109 (2021).

[79] H. Niu, S. Niu, and A. R. Oganov, *Simple and Accurate Model of Fracture Toughness of Solids*, Journal of Applied Physics **125**, 065105 (2019).

[80] P. Ravindran, L. Fast, P. A. Korzhavyi, B. Johansson, J. Wills, and O. Eriksson, *Density Functional Theory for Calculation of Elastic Properties of Orthorhombic Crystals: Application to TiSi$_2$*, Journal of Applied Physics **84**, 4891 (1998).

[81] M. A. Ali, M. M. Hossain, M. M. Uddin, A. K. M. A. Islam, D. Jana, and S. H. Naqib, *DFT Insights into New B-Containing 212 MAX Phases: Hf$_3$AB$_2$ (A = In, Sn)*, Journal of Alloys and Compounds **860**, 158408 (2021).

[82] M. A. Afzal and S. H. Naqib, *A DFT Based First-Principles Investigation of Optoelectronic and Structural Properties of Bi$_2$Te$_2$Se*, Phys. Scr. **96**, 045810 (2021).

[83] B. Rahman Rano, I. M. Syed, and S. H. Naqib, *Elastic, Electronic, Bonding, and Optical Properties of WTe$_2$ Weyl Semimetal: A Comparative Investigation with MoTe$_2$ from First Principles*, Results in Physics **19**, 103639 (2020).

[84] X. Gao, Y. Jiang, R. Zhou, and J. Feng, *Stability and Elastic Properties of Y–C Binary Compounds Investigated by First Principles Calculations*, Journal of Alloys and Compounds **587**, 819 (2014).

[85] C. M. Kube, *Elastic Anisotropy of Crystals*, AIP Advances **6**, 095209 (2016).

[86] V. Arsigny, P. Fillard, X. Pennec, and N. Ayache, *Fast and Simple Calculus on Tensors in the Log-Euclidean Framework*, in *Medical Image Computing and Computer-Assisted Intervention – MICCAI 2005*, edited by J. S. Duncan and G. Gerig, Vol. 3749 (Springer, Berlin, Heidelberg, 2005), pp. 115–122.

[87] C. M. Kube and M. de Jong, *Elastic Constants of Polycrystals with Generally Anisotropic Crystals*, Journal of Applied Physics **120**, 165105 (2016).

[88] S. I. Ranganathan and M. Ostoja-Starzewski, *Universal Elastic Anisotropy Index*, Phys. Rev. Lett. **101**, 055504 (2008).

[89] Y. H. Duan, Y. Sun, M. J. Peng, and S. G. Zhou, *Anisotropic Elastic Properties of the Ca–Pb Compounds*, Journal of Alloys and Compounds **595**, 14 (2014).

[90] D. H. Chung and W. R. Buessem, *The Elastic Anisotropy of Crystals*, Journal of Applied Physics **38**, 2010 (1967).

[91] R. Gaillac, P. Pullumbi, and F.-X. Coudert, *ELATE: An Open-Source Online Application for Analysis and Visualization of Elastic Tensors*, Journal of Physics: Condensed Matter **28**, (2016).

[92] E. N. Koukaras, G. Kalosakas, C. Galiotis, and K. Papagelis, *Phonon Properties of Graphene Derived from Molecular Dynamics Simulations*, Sci Rep **5**, 12923 (2015).

[93] Y. Yun, D. Legut, and P. M. Oppeneer, *Phonon Spectrum, Thermal Expansion and Heat Capacity of UO$_2$ from First-Principles*, Journal of Nuclear Materials **426**, 109 (2012).

[94] G. Kresse, J. Furthmüller, and J. Hafner, *Ab Initio Force Constant Approach to Phonon Dispersion Relations of Diamond and Graphite*, Europhys. Lett. **32**, 729 (1995).

[95] K. Parlinski, Z. Q. Li, and Y. Kawazoe, *First-Principles Determination of the Soft Mode in Cubic ZrO$_2$*, Physical Review Letters **78**, 4063 (1997).

[96] K. Boudiaf, A. Bouhemadou, Y. Al-Douri, R. Khenata, S. Bin-Omran, and N. Guechi, *Electronic and Thermoelectric Properties of the Layered BaFAgCh (Ch = S, Se and Te): First-Principles Study*, Journal of Alloys and Compounds **759**, 32 (2018).

[97] A. Bekhti-Siad, K. Bettine, D. P. Rai, Y. Al-Douri, X. Wang, R. Khenata, A. Bouhemadou, and C. H. Voon, *Electronic, Optical and Thermoelectric Investigations of Zintl Phase $AE_3AlAs_3$ (AE = Sr, Ba): First-Principles Calculations*, Chinese Journal of Physics **56**, 870 (2018).

[98] A. Belhachemi, H. Abid, Y. Al-Douri, M. Sehil, A. Bouhemadou, and M. Ameri, *First-Principles Calculations to Investigate the Structural, Electronic and Optical Properties of $Zn_{1-x}Mg_xTe$ Ternary Alloys*, Chinese Journal of Physics **55**, 1018 (2017).

[99] A. H. Reshak and S. Auluck, *Theoretical Investigation of the Electronic and Optical Properties of $ZrX_2$ (X=S, Se and Te)*, Physica B: Condensed Matter **353**, 230 (2004).

[100] F. Parvin and S. H. Naqib, *Pressure Dependence of Structural, Elastic, Electronic, Thermodynamic, and Optical Properties of van Der Waals-Type $NaSn_2P_2$ Pnictide Superconductor: Insights from DFT Study*, Results in Physics **21**, 103848 (2021).

[101] J.-H. Xu, T. Oguchi, and A. J. Freeman, *Crystal Structure, Phase Stability, and Magnetism in $Ni_3V$*, Phys. Rev. B **35**, 6940 (1987).

[102] H. Hou, Z. Wen, Y. Zhao, L. Fu, N. Wang, and P. Han, *First-Principles Investigations on Structural, Elastic, Thermodynamic and Electronic Properties of $Ni_3X$ (X = Al, Ga and Ge) under Pressure*, Intermetallics **44**, 110 (2014).

[103] W.-C. Hu, Y. Liu, D.-J. Li, X.-Q. Zeng, and C.-S. Xu, *First-Principles Study of Structural and Electronic Properties of C14-Type Laves Phase $Al_2Zr$ and $Al_2Hf$*, Computational Materials Science **83**, 27 (2014).

[104] N. Sadat Khan, B. Rahman Rano, I. M. Syed, R. S. Islam, and S. H. Naqib, *First-Principles Prediction of Pressure Dependent Mechanical, Electronic, Optical, and Superconducting State Properties of $NaC_6$: A Potential High-$T_c$ Superconductor*, Results in Physics **33**, 105182 (2022).

[105] A. K. M. A. Islam and S. H. Naqib, *Possible explanation of high-$T_c$ in some 2D cuprate superconductors*, J. Phys. Chem. Solids **58**, 1153 (1997).

[106] M. M. Mridha and S. H. Naqib, *Pressure dependent elastic, electronic, superconducting, and optical properties of ternary barium phosphides ($BaM_2P_2$; M = Ni, Rh): DFT based insights,* Physica Scripta **95** (10), 105809 (2020).

[107] F. L. Hirshfeld, *Bonded-Atom Fragments for Describing Molecular Charge Densities*, Theoret. Chim. Acta **44**, 129 (1977).

[108] R. C. D. Richardson, *The Wear of Metals by Hard Abrasives*, Wear **10**, 291 (1967).

[109] V. V. Brazhkin, A. G. Lyapin, and R. J. Hemley, *Harder than Diamond: Dreams and Reality*, Philosophical Magazine A **82**, 231 (2002).

[110] F. Gao, *Theoretical Model of Intrinsic Hardness*, Phys. Rev. B **73**, 132104 (2006).

[111] F. Birch, *Finite Strain Isotherm and Velocities for Single-Crystal and Polycrystalline NaCl at High Pressures and 300°K*, J. Geophys. Res. **83**, 1257 (1978).

[112] F. Gao, J. He, E. Wu, S. Liu, D. Yu, D. Li, S. Zhang, and Y. Tian, *Hardness of Covalent Crystals*, Phys. Rev. Lett. **91**, 015502 (2003).

[113] R. D. Harcourt, *Diatomic Antibonding $\Sigma^*s$ Orbitals as "metallic Orbitals" for Electron Conduction in Alkali Metals*, J. Phys. B: Atom. Mol. Phys. **7**, L41 (1974).

[114] J. H. Westbrook and H. Conrad, editors, *The Science of Hardness Testing and Its Research Applications: Based on Papers Presented at a Symposium of the American Society for Metals,* (American Society for Metals, 1973).

[115] J. R. Christman, *Fundamentals of Solid State Physics* (Wiley, New York, 1988).

[116] E. Schreiber, O. L. Anderson, and N. Soga, *Elastic Constants and Their Measurement* (McGraw-Hill, New York, 1974).

[117] O. L. Anderson, *A Simplified Method for Calculating the Debye Temperature from Elastic Constants*, Journal of Physics and Chemistry of Solids **24**, 909 (1963).

[118] G. A. Slack, *The Thermal Conductivity of Nonmetallic Crystals*, in *Solid State Physics*, Vol. 34 (Elsevier, 1979), pp. 1–71.

[119] C. L. Julian, *Theory of Heat Conduction in Rare-Gas Crystals*, Phys. Rev. **137**, A128 (1965).

[120] D. R. Clarke, *Materials Selection Guidelines for Low Thermal Conductivity Thermal Barrier Coatings*, Surface and Coatings Technology **163–164**, 67 (2003).

[121] D. G. Cahill, S. K. Watson, and R. O. Pohl, *Lower Limit to the Thermal Conductivity of Disordered Crystals*, Phys. Rev. B **46**, 6131 (1992).

[122] M. E. Fine, L. D. Brown, and H. L. Marcus, *Elastic Constants versus Melting Temperature in Metals*, Scripta Metallurgica **18**, 951 (1984).

[123] M. I. Naher and S. H. Naqib, *An Ab-Initio Study on Structural, Elastic, Electronic, Bonding, Thermal, and Optical Properties of Topological Weyl Semimetal TaX (X = P, As)*, Sci Rep **11**, 1 (2021).

[124] M. F. Ashby, P. J. Ferreira, and D. L. Schodek, *Material Classes, Structure, and Properties*, in *Nanomaterials, Nanotechnologies and Design* (Elsevier, 2009), pp. 87–146.

[125] M. I. Naher and S. H. Naqib, *A Comprehensive Study of the Thermophysical and Optoelectronic Properties of $Nb_2P_5$ via Ab-Initio Technique*, Results in Physics **28**, 104623 (2021).

[126] M. S. Dresselhaus, *SOLID STATE PHYSICS PART II Optical Properties of Solids* (Citeseer, 1999).

[127] M. A. Ali, M. M. Hossain, A. K. M. A. Islam, and S. H. Naqib, *Ternary boride $Hf_3PB_4$: Insights into the physical properties of the hardest possible boride MAX phase*, Journal of Alloys and Compounds **857**, 158264 (2021).

[128] M. A. Ali, M. M. Hossain, M. M. Uddin, M. A. Ali, A. K. M. A. Islam, and S. H. Naqib, *Physical properties of new MAX phase borides $M_2SB$ (M= Zr, Hf and Nb) in comparison with conventional MAX phase carbides $M_2SC$ (M= Zr, Hf and Nb): Comprehensive insights*, Journal of Materials Research and Technology **11**, 1000 (2021).

[129] M. A. Ali, M. A. Hadi, M. M. Hossain, S. H. Naqib, and A. K. M. A. Islam, *Theoretical investigation of structural, elastic, and electronic properties of ternary boride MoAlB*, Physica status solidi (b) **254** (7), 1700010 (2017).

**CRediT author statement**